\documentclass{aa}  
\usepackage{graphicx}
\usepackage{graphics}
\usepackage[varg]{txfonts}
\usepackage[colorlinks=true,allcolors=blue]{hyperref} 
\usepackage{enumerate}
\usepackage{amsmath}
\usepackage{natbib}
\usepackage{comment}
\usepackage[para]{threeparttable}
\usepackage{xcolor}
\begin{document}

   \title{Star-forming early- and quiescent late-type galaxies in the local Universe}

   \subtitle{}

   \author{
            E.-D. Paspaliaris\inst{1,2}\and
            E. M. Xilouris\inst{1}\and
            A. Nersesian\inst{3}\and
            S. Bianchi\inst{4}\and
            I. Georgantopoulos\inst{1}\and
            V. A. Masoura\inst{5,1}\and
            \\G. E. Magdis\inst{6,7,8}\and
            M. Plionis\inst{1,2}
            }

   \institute{National Observatory of Athens, Institute for Astronomy, Astrophysics, Space Applications and Remote Sensing, Ioannou Metaxa and Vasileos Pavlou, 15236 Athens, Greece\\
   email: \href{mailto:epaspal@noa.gr}{epaspal@noa.gr}
   \and
   Department of Astrophysics, Astronomy \& Mechanics, School of Physics, Aristotle University of Thessaloniki, 54124 Thessaloniki, Greece
   \and
   Sterrenkundig Observatorium, Universiteit Gent, Krijgslaan 281 S9, 9000 Gent, Belgium
   \and
   INAF - Osservatorio Astrofisico di Arcetri, Largo E. Fermi 5, 50125, Florence, Italy
   \and
   Instituto de Fisica de Cantabria (CSIC-Universidad de Cantabria), Avenida de los Castros, 39005 Santander, Spain
   \and
   Cosmic Dawn Center (DAWN), Jagtvej 128, DK2200 Copenhagen N, Denmark
   \and
   DTU-Space, Technical University of Denmark, Elektrovej 327, DK2800 Kgs. Lyngby, Denmark
   \and
   Niels Bohr Institute, University of Copenhagen, Blegdamsvej 17, DK2100 Copenhagen Ø, Denmark
   }
   \date{Received Aug 23, 2022 / Accepted Sep 27, 2022}

 
  \abstract
   {}
   {The general consensus is that late-type galaxies undergo intense star-formation activity while early-type galaxies are mostly inactive. We question this general rule and investigate the existence of star-forming early-type and quenched late-type galaxies in the local universe. By computing the physical properties of these galaxies and by using information on their structural properties as well as the density of their local environment we seek for understanding the differences from their `typical' counterparts.}
   {We make use of the multi-wavelength photometric data (from the ultra-violet to the sub-millimetre), for 2,209 morphologically classified galaxies in the Galaxy And Mass Assembly survey. Furthermore, we separate the galaxies into subsets of star-forming and quiescent based on their dominant ionising process, making use of established criteria based on the $W_{\text{H}\alpha}$ width and the [N$_\text{II}$/H$\alpha$] ratio. Taking advantage of the spectral energy distribution fitting code \texttt{CIGALE} we derive galaxy properties, such as the stellar mass, the dust mass, and the star formation rate and also estimate the unattenuated and the dust-absorbed stellar emission, for both the young ($\leq$ 200 Myr) and old (> 200 Myr) stellar populations. 
   }
   {We find that about 47\% of E/S0 galaxies in our sample show ongoing star-formation activity and 8\% of late-type galaxies are quiescent. The star-forming elliptical galaxies, together with the little blue spheroids, constitute a population that follows very well the star-forming main-sequence of spiral galaxies. The fraction of the luminosity originating from young stars in the star-forming early-type galaxies is quite substantial ($\sim25\%$) and similar to that of the star-forming late-type galaxies. The stellar luminosity absorbed by the dust (and used to heat the dust grains) is highest in star-forming E/S0 galaxies (an average of 35\%) followed by star-forming Sa-Scd galaxies (27\%) with this fraction becoming significantly smaller for their quiescent analogues (6\% and 16\%, for E/S0 and Sa-Scd, respectively). Star-forming and quiescent E/S0 galaxies donate quite different fractions of their young stellar luminosities to heat up the dust grains (74\% and 36\%, respectively) while this fractions are very similar for star-forming and quiescent Sa-Scd galaxies (59\% and 60\%, respectively). Investigating possible differences between star-forming and quiescent galaxies we find that the intrinsic (unattenuated) shape of the SED of the star-forming galaxies is, on average, very similar for all morphological types. Concerning their structural parameters, quiescent galaxies tend to show larger values of the r-band S\'{e}rsic index and larger effective radii (compared to star-forming galaxies). Finally, we find that star-forming galaxies preferably reside in lower-density environments compared to the quiescent ones, which exhibit a higher percentage of sources being members of groups.}
   {}


   \maketitle
%


\section{Introduction}

There is a general perception that the star-forming activity in galaxies is strongly correlated to their morphological type.
Gas-poor elliptical (E), lenticular (S0) and dwarf galaxies form stars in rates significantly lower than $\sim$1 M$_{\odot}$ yr$^{-1}$, while gas-rich spirals and irregulars (Irr) have star-formation rates (SFRs) that can reach up to $\sim$20 M$_\odot$ yr$^{-1}$ (\citealt{Kennicutt1983}; \citealt{Gao2004}; \citealt{Calvi2018}; \citealt{Nersesian2019b}). Much higher SFRs, exceeding several hundreds of M$_\odot$ yr$^{-1}$, can be found in local starburst galaxies and ultra/luminous infrared galaxies (U/LIRGs) (\citealt{daCunha2010}; \citealt{Combes2013}; \citealt{Kennicutt2021}; \citealt{Paspaliaris2021}). 

E galaxies are considered to be amongst the most massive, old and red systems (\citealt{Bernardi2003}; \citealt{Kelvin2014b}; \citealt{GonzalezDelgado2015}; \citealt{Nersesian2019b}). It is believed that they formed either by the collapse of protogalaxies, having a prominent early burst of star formation and then evolve passively (\textit{monolithic} view; \citealt{PartridgePeebles1967}; \citealt{Larson1975}) or by merging galaxies with an unclear evolutionary path following (\textit{hierarchical} view; \citealt{Toomre1972}), leading to the quiescent (Q) systems we observe today. In most cases, their stellar content is highly concentrated in the centre with its density decreasing towards the outskirts of the galaxy. The interstellar medium (ISM; if any) is mostly concentrated in the galaxy centre. Conversely, late-type spiral galaxies are mostly bluer, actively star forming systems, with a central bulge consisting mainly of old stars, with ongoing star-formation activity occurring in the dusty spiral arms. Furthermore, the existence of a bar can lead to the funnelling of gas towards the galactic centre playing a significant role in the evolution of the properties of the host galaxy (e.g. \citealt{Sorensen1976}; \citealt{Athanassoula2013}, and references therein). A hybrid-like case also exists consisting of a blue, low-mass, compact spheroidal population often referred to as "Little Blue Spheroids" (LBSs). This population of galaxies are structurally similar to their higher-mass elliptical galaxy analogues, but with the scaling relations of their physical properties, such as SFR, stellar mass ($M_\text{star}$), and bolometric luminosity resembling star-forming (SF) spiral galaxies (\citealt{Mahajan2015}; \citealt{Mahajan2018}). Moreover, they lie outside the standard Hubble parametrisation range \citep[$T$;][]{Makarov2014} which is defined to include galaxies with Hubble stage from $T$=-5 (pure Ellipticals) to $T$=10 (Irregulars).
   
Although the trend in the average SFRs with the morphology is strong, SFRs of galaxies of the same type may exhibit a dispersion of 1 dex \citep{Kennicutt1998}. Additionally, a bimodal distribution has been found for the SFR of galaxies \citep[e.g.][]{Wetzel2012,Trussler2020,Kalinova2021,Sampaio2022}. Combining two fundamental galaxy properties, the SFR and $M_\text{star}$, we can quarry information about their current rate of conversion of the gas into stars. In the case of SF galaxies, the two parameters are found to be tightly correlated, occupying a distinct region in the SFR-$M_\text{star}$ diagram, often referred to as the `star-forming main sequence' (SFMS) of galaxies (e.g. \citealt{Noeske2007}; \citealt{Elbaz2007}; \citealt{Wuyts2011}; \citealt{Whitaker2012}, and references therein) or `the blue cloud'. On the contrary, Q galaxies exhibit a weaker relation between SFR and $M_\text{star}$, occupying the area below the SFMS, forming the `red cloud'. The `blue cloud' consists mainly of late-type galaxies (Sa-Irr; hereafter LTGs), while the `red cloud' is mostly occupied by early-type galaxies (E and S0; hereafter ETGs\footnote{Due to their spheroidal morphology LBS galaxies are also considered as ETGs in the current study.}). This bimodality has been thoroughly investigated in previous studies, such as \citet{Strateva2001}, \citet{Blanton2003}, \citet{Baldry2004} and \citet{Taylor2015}. However, several studies have reported the existence of ETGs with on-going star formation and also of LTGs with ceased star-forming activity (e.g. \citealt{Rowlands2012}; \citealt{Bitsakis2019}; \citealt{CanoDiaz2019}). 

It has been shown that the rate that galaxies form stars can be strongly influenced by the environment they reside in (e.g \citealt{Bersanti2018}; \citealt{Davies2019}; \citealt{Sampaio2022}). When spiral galaxies traverse the dense inter-cluster medium, their interstellar gas is removed through ram-pressure stripping and they lose their ability to form new stars (e.g. \citealt{GunnGott1972}; \citealt{Dressler1980}). In this manner, groups and clusters mainly consist of elliptical and gas-poor galaxies which have different properties from their counterparts settled in less dense environments (e.g. \citealt{Baldry2006}; \citealt{SkibbaSheth2009}). Other proposed mechanisms through which galaxies may lose their ability to form new stars or even perturb their morphology are strangulation (\citealt{Larson1980}; \citealt{Kauffmann1993}; \citealt{Diaferio2001}), harassment \citep{Moore1996} and minor-mergers of tidal interactions \citep{Park2008}. \citet{McIntosh2014} and \citet{Haines2015} suggested that SF Es have suffered a recent morphological transition without having enough time to exhaust their gas reservoir, being in a post-starburst phase. Alternatively, Es in low density environments may be rejuvenated by eventually accreting cold gas as suggested by \citet{Thomas2010}.

Spectral energy distribution (SED) modelling techniques allow us to decompose the SED of galaxies and to derive useful information about their different emitting components. These techniques usually combine the stellar emission (originating from both old and young stars) in ultra-violet (UV), optical and near-infrared (NIR) wavelengths with the dust infrared (IR) emission so that the energy budget is fully conserved. In addition, a star-formation activity through cosmic time (i.e. the star-formation history; SFH) is assumed allowing for the determination of the current SFR but also the buildup of the current stellar mass of the galaxy. Having set a grid of input parameters the SED-fitting codes can then retrieve the best set of templates that better reproduce the observations and, thus, provide useful information about physical properties of the galaxies, such as, the current SFR and the stellar and the dust masses as well as luminosities of the different components of the stellar populations and the dust.

   \begin{figure*}[t!]
   \centering
   \includegraphics[width=\textwidth]{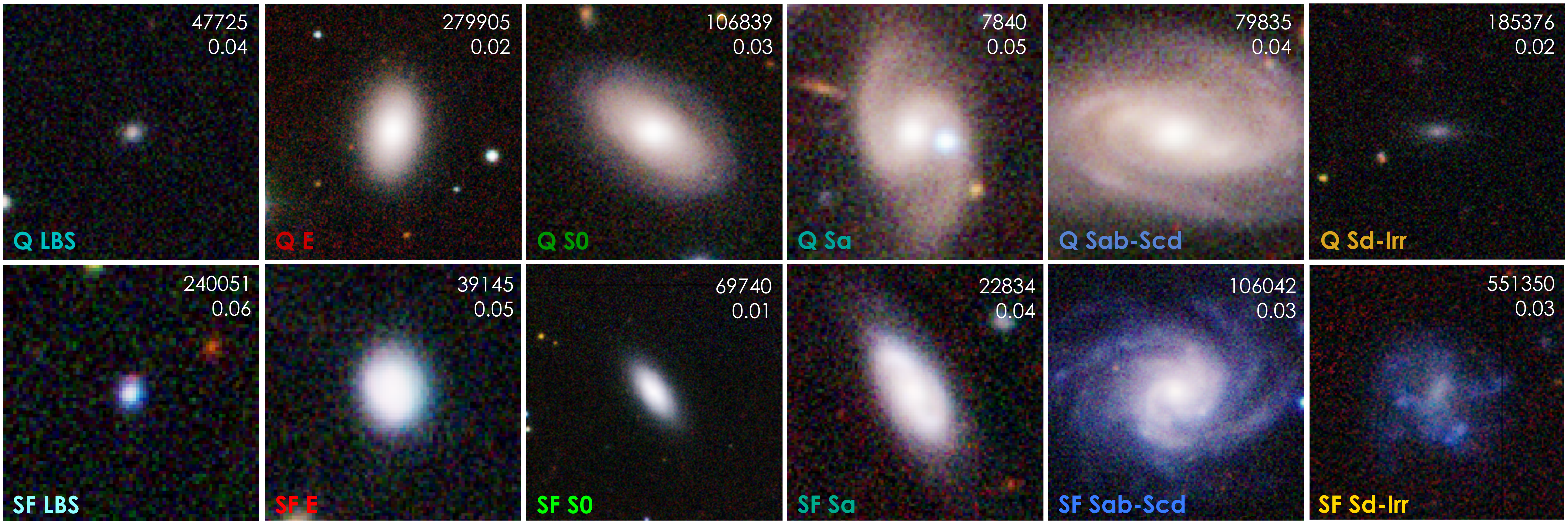}
   \caption{Typical GAMA galaxies with different morphologies (LBS, E, S0, Sa, Sab-Scd, Sd-Irr; from left to right) and different star-forming activity according to their dominant ionisation process (Q, SF; top and bottom respectively). A composite g-, i-, H-band image is shown for each galaxy in a frame of 30 kpc $\times$ 30 kpc. The GAMA catalogue identification number and the redshift of each galaxy is indicated at the top-right of each panel.}
   \label{fig:giH}
   \end{figure*}

In this paper we perform SED modelling of 2,209 local \mbox{(z $<0.06$)} galaxies in the Galaxy And Mass Assembly (GAMA) survey\footnote{http://www.gama-survey.org} using the \texttt{CIGALE}\footnote{https://cigale.lam.fr} SED fitting code. For all the galaxies there is available morphological classification as well as classification according to their star-forming activity as traced by optical spectral lines (available in the GAMA survey). We examine all morphological types separately, but, mainly, focusing on SF ETGs and Q LTGs. These subsets constitute obvious exceptions of the general rule having ETGs being, mostly, quenched and LTGs, mostly, actively forming new stars. The scope of this study is to explore the physical and structural parameters that make them differentiate from their `typical' counterparts. In Sect.~\ref{sec:sample} we present the sample properties and classifications, as well as a description of the SED-fitting method used. In Sect.~\ref{sec:properties} we present the typical SEDs as a function of morphology and star-forming activity as well as the physical properties of the SF ETGs and Q LTGs. The different stellar populations in different types of galaxies as well as their role in the heating of the dust is discussed in Sect.~\ref{sec:populations}. A discussion follows in Sect.~\ref{discussion} describing the basic differences that we find between SF and Q galaxies, while our findings are summarised in Sect.~\ref{sec:sum}. A comparison with the results provided in the GAMA survey obtained by the SED-fitting code \texttt{MAGPHYS} is presented in Appendix \ref{ap:validation}. Throughout the paper, we adopt $H_0$ = 70 km s$^{-1}$ Mpc$^{-1}$, $\Omega$$_\text{m}$ = 0.3,  $\Omega_\Lambda$ = 0.7, and a \citet{Salpeter} initial mass function (IMF).


\section{Data and analysis}\label{sec:sample}

\subsection{Sample selection}

The multi-wavelength data used in this work are provided by the Panchromatic Data Release \citep[from FUV to FIR;][]{Driver2016} of the GAMA survey \citep{Driver2009, Driver2011}. For each source, there are available observations for up to 21 different broad-band filters from five observatories. All the filters provided by the GAMA survey, and used in the current analysis, are listed in Table~\ref{tab:bands} (see also Table 1 in \citealt{Wright2016}). The fluxes of the galaxies in the different bands were calculated using the Lambda Adaptive Multi-Band Deblending Algorithm in R (LAMBDAR\footnote{https://github.com/AngusWright/LAMBDAR}; \citealt{Wright2016}). LAMBDAR performs deblended aperture-matched photometry without requiring PSF or pixel matched images. Four major processes are employed, aperture convolution, aperture deblending, sky subtraction and aperture correction accounting for the different PSFs. Hence, the algorithm avoids the necessity of the seeing and imaging being matched and in contrast to other algorithms the images do not need to be smoothed to the resolution of the lowest quality band. Fluxes, in the FUV-K wavelength range, were also corrected for Galactic extinction.

   \begin{table}
   \caption{List of the 21 broad-band filters provided by the GAMA database and used in the current SED-fitting analysis.}
   \label{tab:bands}
   \tiny
   \begin{center}
   \begin{tabular}{l c}
\hline
\hline
   \multicolumn{1}{l}{Observatory} &
   \multicolumn{1}{c}{Band (central wavelength)} \\
\hline
   GALEX & FUV (1550\AA), NUV (2275\AA) \\
   SDSS & u (3540\AA), g (4770\AA), r (6230\AA), i (7630\AA), z (9134\AA) \\
   VISTA & Z (8770\AA), Y (1.0$\mu$m), J (1.3$\mu$m), H (1.7$\mu$m), K (2.2$\mu$m) \\
   WISE & W1 (3.4$\mu$m), W2 (4.6$\mu$m), W3 (12$\mu$m), W4 (22$\mu$m) \\
   HSO & PACS (100$\mu$m, 160$\mu$m) , SPIRE (250$\mu$m, 350$\mu$m, 500$\mu$m) \\
\hline
   \end{tabular}
   \end{center}
   \end{table}


From the full sample of the GAMA survey we have selected a subsample of galaxies in the GAMA II equatorial survey regions. All galaxies in this subsample have been visually classified according to their Hubble type using UKIDSS H and SDSS i and g band images as described in detail in \citet{Kelvin2014a} and \citet{Moffett2016}. The classification was held in two phases. In the first phase logarithmically scaled (in brightness) 20" $\times$ 20" images were used, while in the second phase 40" $\times$ 40" images scaled with the arctan function (in brightness) were visually inspected. The arctan function was chosen as a more effective method against manually defined upper or lower brightness levels, which, potentially, lead to misclassification of galaxies.
A decision tree with three levels of classification options was followed by three independent pairs of observers. In the first level the observers had to divide the sample by type, into Spheroid Dominated, Disk Dominated, LBS or Stars (in the case that a foreground star is in front of the primary object or a supernova exists within a galaxy, the primary object is classified as a star and is subsequently rejected). The options at the other two classification levels were Single/Multi Component and Barred/Unbarred. The final classification was determined by the majority opinion. This classification has been used in various studies, so far, such as \citet{Kelvin2014b, Kelvin2018}, \citet{Agius2013}, \citet{Moffett2019}, \citet{Mahajan2020} and \citet{Bellstedt2020}.

This volume- and luminosity-limited sample is referred to as GAMAnear and consists of 6,433 galaxies in the redshift range \mbox{0.002 < $z$ < 0.06} and an extinction corrected r-band SDSS Petrosian magnitude of $r$ < 19.8 mag. For the z<0.02 galaxies, the flow-corrected redshifts (using the \citealt{Tonry2000} flow model) were used (see \citealt{Baldry2012}). After applying two additional selection criteria that all galaxies are observed by the \textit{Herschel Space Observatory} \citep[HSO;][]{Pilbratt2010}, and, also, have available optical spectroscopic information (with the available optical lines detected at S/N>3), we ended up with 2,597 galaxies. The selection of HSO observed sources allows us to accurately compute the dust mass, while the availability of optical spectroscopic indices provides information on the star formation and nuclear activity for each galaxy. Further structural measurements, such as the \mbox{effective radius ($R_\text{e}$)} and S\'{e}rsic index for all the galaxies in our sample were provided by \cite{Kelvin2012}, while information of the density of the local environment for 1,221 galaxies is given in \citet{Robotham2011}. 

Examples of different Hubble-type galaxies with different star-forming activity (according to optical emission lines; see Sect.~\ref{sec:classification}) are shown in Fig.~\ref{fig:giH}. It is worth noticing that even a visual inspection reveals a strong difference in the appearance of these galaxies with Q galaxies (top panels) having redder colours compared to their bluer SF counterparts (bottom panels). After applying the aforementioned criteria, but also excluding galaxies with possible contamination from an active galactic nucleus (AGN) activity (see Sect.~\ref{sec:classification}), the final sample consists of 269 E, 251 S0-Sa, 659 Sab-Scd, 704 Sd-Irr galaxies and 326 LBS. For the purposes of the current study we performed a more detailed classification of the S0-Sa sources into two different groups (S0 and Sa). This classification was conducted by three of the authors of this study by inspecting images using the GAMA Single Object Viewer and the SDSS SkyServer. Sources exhibiting spiral arms were classified as Sa while the rest were considered S0 galaxies. The observers independently classified the 251 sources and the final classification was assigned by majority agreement, resulting in 224 S0 and 27 Sa galaxies (see Table~\ref{tab:sample}).

As a reference sample for the very local galaxies, we use the dataset of the DustPedia project\footnote{http://dustpedia.astro.noa.gr}. DustPedia includes FUV to submm data of 875 nearby galaxies ($D_\text{L}$ < 40 Mpc), all observed by HSO. The Hubble stage $T$ is available as a morphology indicator for the full sample. For more information about the DustPedia project, we refer the reader to \citet{Davies2019} and \citet{Clark2018}.  For 814 galaxies \citet{Nersesian2019b} performed SED-fitting analysis, in exactly the same way described in this work (see Sect.~\ref{sec:fitting}), delivering their physical properties.

\subsection{Classification by ionisation processes and star-forming activity}\label{sec:classification}

   \begin{figure}[t!]
   \centering
   \includegraphics[width=0.5\textwidth]{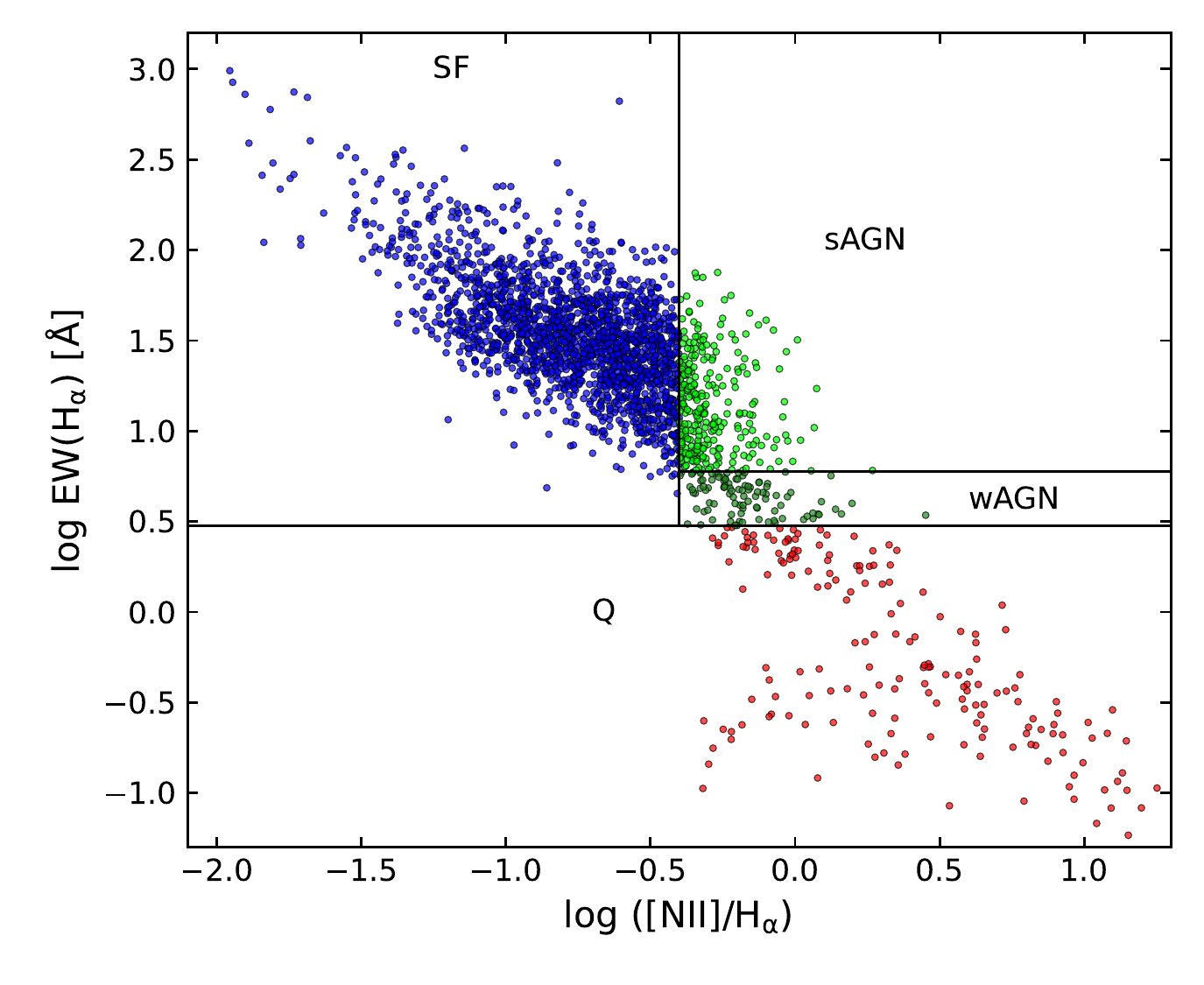}
   \caption{WHAN diagram for the classification of the galaxies due to their dominant ionisation processes. Demarcation lines are from \citet{CidFernandes2011}, for star-forming and quiescent galaxies (blue and red circles, respectively) and strong and weak AGN (light and dark green circles, respectively).}
   \label{fig:whan}
   \end{figure}

Since GAMA is not only a photometric but also a spectroscopic survey \citep{Driver2009,Hopkins2013,Gordon2017} we are able to classify the galaxies in our sample, not only by their morphological characteristics, but also by their dominant ionisation processes. Emission line diagnostic diagrams (e.g. \citealt{BPT1981}; \citealt{Kauffmann2003}; \citealt{Kewley2006}) are widely used to probe the ionization sources in galaxies (e.g., star-formation, AGN, emission by evolved stellar populations). For such a categorisation in this study we make use of the available measurements of the $\text{H}\alpha$ line width ($W_{\text{H}\alpha}$) and the flux ratio of [N$_{\text{II}}$]/H$\alpha$. These quantities are physically independent with $W_{\text{H}\alpha}$ being a measure of the ionising photons absorbed by the gas, relative to the stellar mass, while [N$_{\text{II}}$]/H$\alpha$ a measure of the dependency of the ionisation state and the gas temperature on the nitrogen gas abundance. The topology of these quantities define areas of low- and high-ionisation as well as of star-forming and quiescence stages and it is widely known as the WHAN diagram \citep{CidFernandes2010,CidFernandes2011}. In contrast to other similar classification schemes, WHAN diagram requires only two emission lines (H$\alpha$ and [N$_{\text{II}}$]$\lambda$6584) which, in fact, are two of the most prominent and easy to measure lines in the optical spectra of galaxies. This permits us to maximise the number of sources in the original sample with sufficient spectroscopic data and, thus, to infer the relevant information. The emission lines in the optical spectra of the GAMA sources were fitted using \texttt{GANDALF} \citep{Sarzi2006} and a diffuse (stellar continuum) obscuration correction, caused by diffuse dust in the galaxy, was applied using a \citet{Calzetti2001} obscuration curve (see \citealt{Hopkins2013} for a detailed description of the GAMA spectroscopic analysis).

We present our physically motivated classification scheme in a WHAN diagram in Fig.~\ref{fig:whan}. Within the WHAN diagram we define four classes of galaxies [Q, SF, strong AGN (sAGN), and weak AGN (wAGN)]. {\mbox{log$W_{\text{H}\alpha}$ = 0.48}} splits our sample in line-less, Q galaxies  (with \mbox{log$W_{\text{H}\alpha}$ < 0.48}) from emission line galaxies (ELGs; with \mbox{log$W_{\text{H}\alpha} \geq 0.48$}). 
In the area of ELGs the vertical line at log[N$_{\text{II}}$]/H$\alpha$ = -0.4 distinguishes the sources where star formation is responsible for the ionising photon output from those that their spectrum can only be explained by a harder ionising field originating from an AGN. Sources with \mbox{log$W_{\text{H}\alpha}$ $\geq$ 0.48} and \mbox{log[N$_{\text{II}}$]/H$\alpha$ < -0.4} are classified as pure SF galaxies. The AGN locus consists of two areas; sAGN occupy the area where log$W_{\text{H}\alpha}$ $\geq$ 0.78 and log[N$_{\text{II}}$]/H$\alpha$ > -0.4, while wAGN are the ones with \mbox{0.48 $\leq$ log$W_{\text{H}\alpha}$ $\leq$ 0.78} and log[N$_{\text{II}}$]/H$\alpha$ $\geq$ -0.4. For the purposes of this work, we do not consider AGNs, so, sources in the sAGN and wAGN regions of this diagram are excluded from the sample of 2,597 ending up with 2,209 galaxies. We caution the reader that out of these galaxies, two sub-classes, Q LBS and Q Sd-Irr are underrepresented (3 and 12 galaxies respectively) so they are not fully considered in the subsequent statistical analysis.
   
   \begin{table}
   \caption{Numbers of galaxies in different morphology bins and star-formation activity. AGN sources have been excluded (see Sect.\ref{sec:classification}).}
   \label{tab:sample}
   \tiny
   \begin{center}
   \begin{tabular}{l||c c|c c|c c}
\hline
\hline
   \multicolumn{1}{l||}{morph. bin} &
   \multicolumn{1}{c}{$N_\text{obj}$} &
   \multicolumn{1}{c|}{perc.\%} &
   \multicolumn{1}{c}{Q$_\text{WHAN}$} &
   \multicolumn{1}{c|}{SF$_\text{WHAN}$} &
   \multicolumn{1}{c}{Q$_\text{sSFR}$} &
   \multicolumn{1}{c}{SF$_\text{sSFR}$} \\
\hline
   LBS & 326 & 14.7 & 3 & 323 & 1 & 325 \\
   E & 269 & 12.2 & 135 & 134 & 141 & 128 \\
   S0 & 224 & 10.1 & 127 & 97 & 123 & 101 \\
   Sa & 27 & 1.2 & 15 & 12 & 9 & 18 \\
   Sab-Scd & 659 & 29.9 & 37 & 622 & 23 & 636 \\
   Sd-Irr & 704 & 31.9 & 12 & 692 & 14 & 690 \\
\hline
   \end{tabular}
   \end{center}
   \end{table}

   
Another way to separate SF and Q galaxies in the local universe, often adopted in the literature (e.g. \citealt{Brinchmann2004, Fontanot2009, Donnari2019, Florez2020}), includes setting a threshold of the specific star-formation rate (sSFR) of $10^{-11}$ yr$^{-1}$. Galaxies with sSFR above this value are defined as SF, while the ones below are Q. Although this method is more empirical and it may depend on the manner that stellar mass and SFR are calculated, it agrees well with the first method. The number of Q and SF galaxies predicted by the two methods are presented in Table~\ref{tab:sample} where a direct comparison can be made.

\subsection{SED modelling}\label{sec:fitting}

In this study we use \texttt{CIGALE} SED fitting code (see \citealt{Boquien2019}, and references therein) to model the SEDs of the galaxies in our sample. Having defined a grid of values for the parameters of the various modules for the stellar, gas, dust emission and taking into account the dust attenuation, we use multi-wavelength observations, of each galaxy, to compare with the library of model SEDs created by \texttt{CIGALE}. \texttt{CIGALE} includes all the different components in such a way so that the amount of energy absorbed and re-emitted by the dust grains is fully conserved \citep{Noll2009, Roehlly2014}. In the final phase of this process, the real observations and the SED libraries created by \texttt{CIGALE} are compared using a Bayesian approach and global properties such as the $M_\text{star}$, the dust mass ($M_\text{dust}$), the SFR, the minimum intensity ($U_\text{min}$) of the Inter-Stellar Radiation Field (ISRF), as well as the different emitting components [old/young stellar population, diffuse/photodissociation regions (PDR) dust emission] are derived.

Since we aim to compare the properties of the galaxies in this sample with the very local galaxies, and for consistency purposes, we adopt the parameter space used in the reference DustPedia sample, introduced by \citet{Nersesian2019b}. In this study, a flexible SFH is used, allowing for a late instantaneous burst or quenching (i.e. module \texttt{`sfhdelayedbq'}; \citealt{Ciesla2015}) while the \citet{Bruzual&Charlot} stellar population model, of solar metallicity, along with the \citet{Salpeter} IMF build the stellar components. Two stellar populations are considered, an old \mbox{($>$ 200 \text{Myr})} and a young ($\leq$ 200 Myr). The emission from the stellar components as well as from the ionised gas surrounding massive stars (i.e. nebular emission) are attenuated using a power-law-modified starburst attenuation curve (i.e. module \texttt{`dustatt\_calzleit';} \citealt{Calzetti2000}), while the \textsc{THEMIS} dust model \citep{Jones2017} accounts for the dust emission parameters. The number of free parameters used in the current analysis is ten while a total of 320,166,000 models were produced. The parameter grid used in the current work can be found in Table 1 in \citet{Nersesian2019b}.

For validation purposes, the \texttt{CIGALE}-derived physical properties are compared with the corresponding estimations using the \texttt{MAGPHYS} fitting code in Appendix~\ref{ap:validation}.

   \begin{figure}[t!]
   \centering
   \vspace{0.2cm}\includegraphics[width=0.5\textwidth]{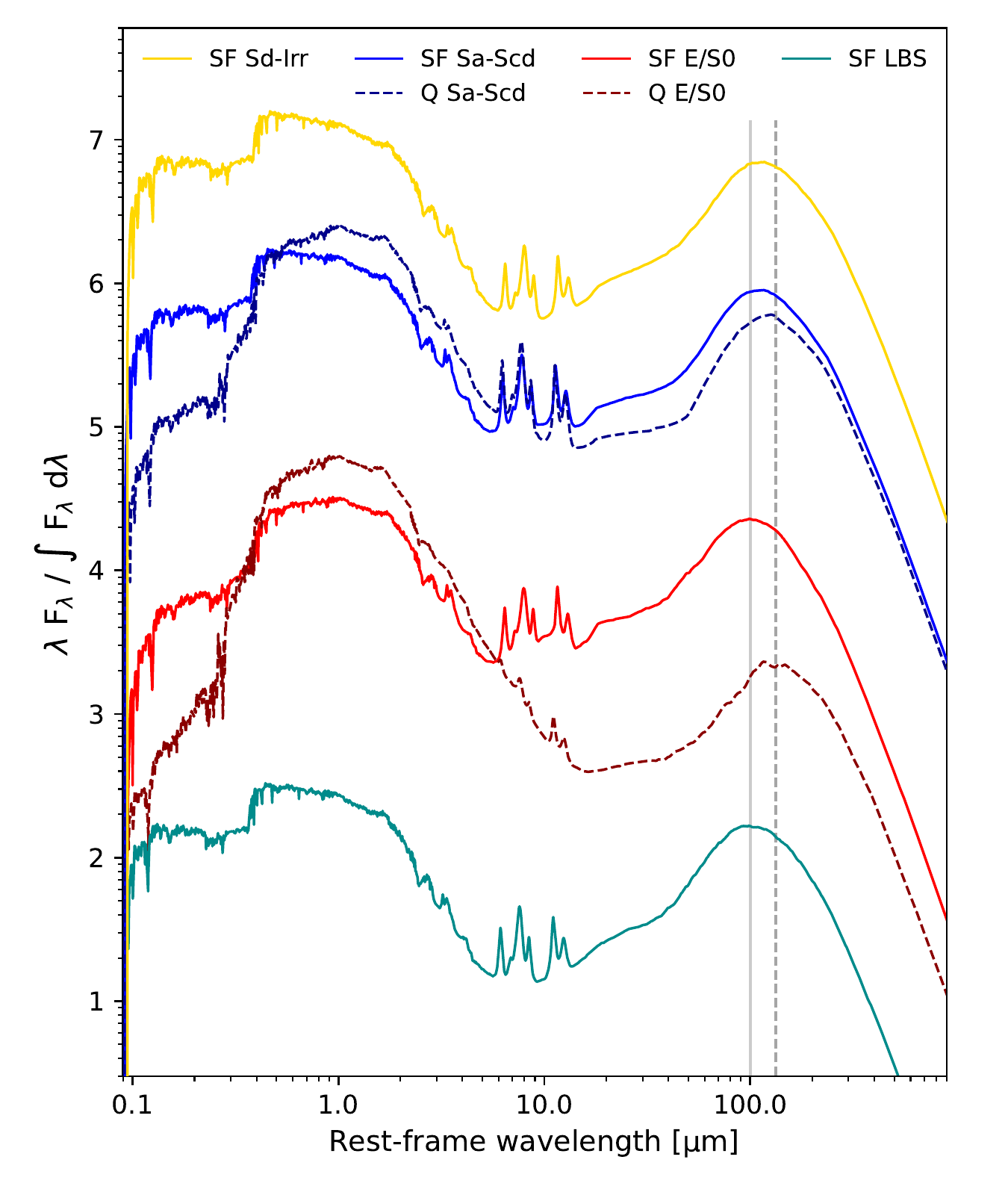}
   \caption{Median template SEDs for different morphological classes shifted arbitrarily for clarity. Solid curves represent the median SEDs of SF galaxies, while dashed SEDs stand for their Q counterparts. The two vertical lines, solid and dashed, connect, approximately, the dust emission peaks in the SEDs of the SF and Q galaxies, respectively.}
   \label{fig:templateSEDs}
   \end{figure}


\section{SEDs and physical properties}\label{sec:properties}

\subsection{Template SEDs} \label{sec:tempSEDs} 

Galaxies of different morphologies have different stellar and dust content usually distributed, within the galaxies, in a very different way. 
These differences are expected to be detectable in their energy output, making their SEDs a unique signature of their morphologies (see \citealt{Ciesla2018}, \citealt{Bianchi2018}, \citealt{Nersesian2019b}, and references therein). Taking advantage of the numerous galaxies in the GAMA sample and the relatively wide redshift range of local galaxies sample (up to redshift 0.06) we present, in Fig.~\ref{fig:templateSEDs}, the median SEDs for the four morphological bins (Sd-Irr, Sa-Scd, E/S0, LBS; top to bottom), separated in subsets according to their major ionising process (SF and Q).

   \begin{figure*}[t!]
   \centering
   \includegraphics[width=\textwidth]{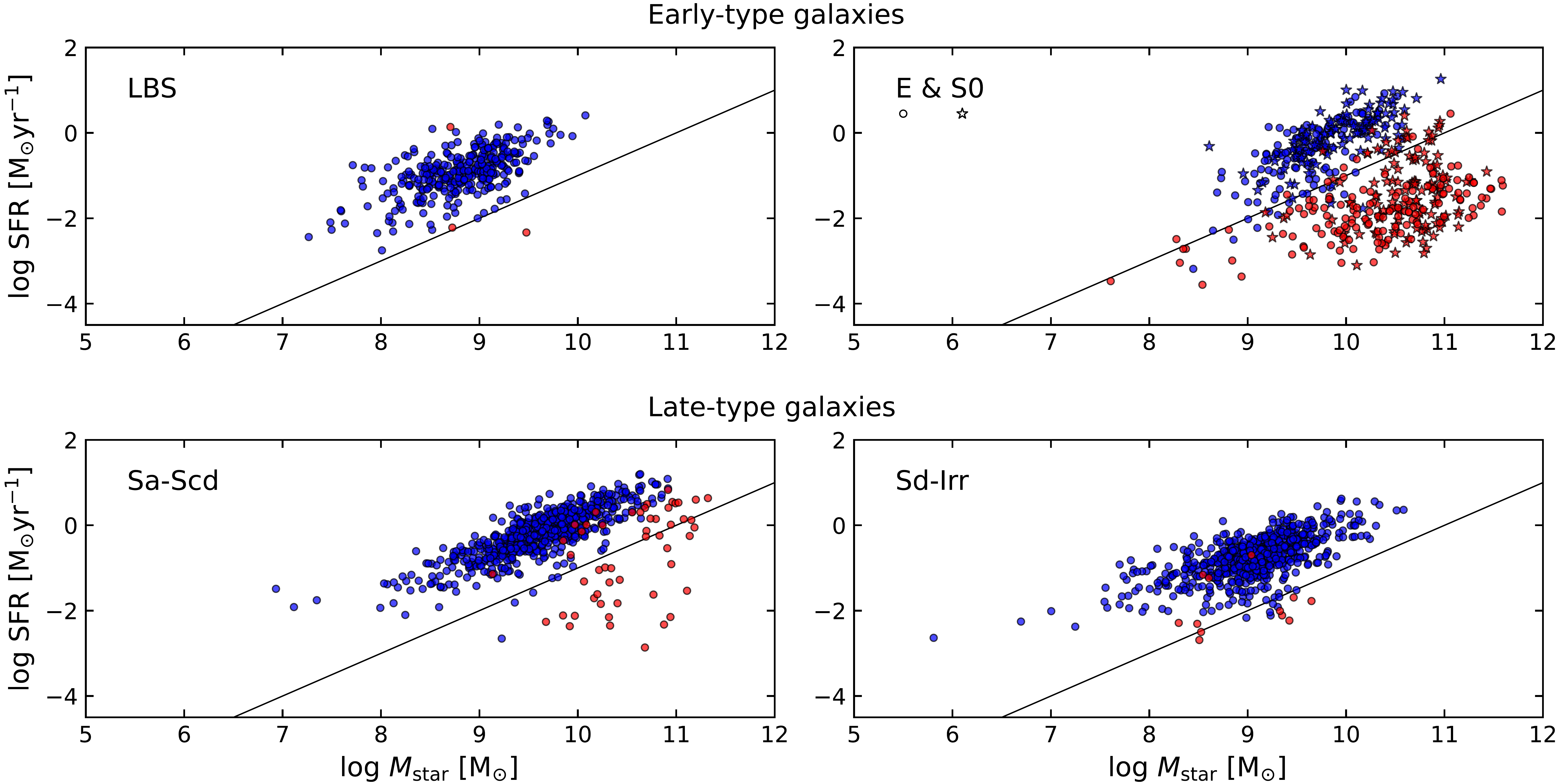}
   \caption{SFR vs $M_\text{star}$ diagram for the subsamples separated by morphological classification. Early-type galaxies are plotted in the \textit{top panels}, while late-type galaxies are plotted in the \textit{bottom panels}. The colouring is same as in Fig.~\ref{fig:whan}. In the \textit{top-right panel} Es are represented by circles, while stars stand for the S0 sources. The sSFR=10$^{-11}$ yr$^{-1}$ line is also indicated, separating the galaxies in SF and Q in quite good agreement with the classification by the WHAN diagram.}
   \label{fig:m-s/morph}
   \end{figure*}
   
After normalising the best fitted, by \texttt{CIGALE}, SED of each galaxy to its bolometric luminosity, we calculated a median SED per morphological bin and per star-forming activity by computing the median specific flux density per wavelength bin. A visual inspection of the shape of the SEDs in Fig.~\ref{fig:templateSEDs} already provides qualitative information about the stellar populations and dust content as well as the star-forming activity for each galaxy type. As stated earlier, since Sd-Irr and LBS lack Q-type galaxies, we do not present the respective SEDs.

The first thing that is obvious is that, the SF galaxies show enhanced dust emission compared to their Q counterparts of the same bolometric luminosity. The measured flux difference at 100 $\mu$m between SF and Q is high in E/S0s (1.1 dex) with lower difference observed in Sa-Scd types (0.21 dex). This indicates that the dust emission may be a clear signature of star-forming activity in some classes of galaxies (E/S0s) but not an obvious one for others (Sa-Scds) where both, SF and Q, types show similar dust SEDs. The stellar emission of the SF galaxies, on the other hand, is always less in NIR and most of the optical ($>0.4 \mu$m) wavelengths compared to Q galaxies of the same bolometric luminosity, but it gets higher, again, below $\sim 0.4 \mu$m.
The peak of the stellar emission also shows clear differences, between SF and Q galaxies of the same bolometric luminosities, with SF galaxies being fainter at 1 $\mu$m by 0.21 and 0.30 dex for Sa-Scd, and E/S0s respectively. This effect is to be expected given the higher dust content observed in SF galaxies.

Another thing that is also obvious is that SF galaxies contain warmer dust compared to the Q galaxies of the same morphological type. This is shown with the peak of the dust emission in the SF galaxies always being to the left of the corresponding peak of the Q galaxies (see the two vertical lines).


\subsection{Galaxy morphology in the SFR-M$_\text{star}$ diagram}\label{subsec:MS_morph}

Despite the slightly different slopes and the scatter around the SFMS found in different studies, mainly depending on the methods used for estimating the SFR as well as the differences in the samples used \citep[][and references therein]{Katsianis2020}, quiescent systems always lie below the SFMS, forming a separate distribution.

In Fig.~\ref{fig:m-s/morph} we present the SFR-$M_\text{star}$ diagram for the galaxies in our sample of different morphological types and star-forming activity.
Independently of the two methods used to classify the galaxies in SF and Q (see Sect.~\ref{sec:classification}), it is obvious that different morphological types have different fractions of SF and Q galaxies with the separation between the two populations being more prominent in the earlier-type galaxies (E/S0; top-right panel). As mentioned earlier (Table~\ref{tab:sample}) Q LBS and Q Sd-Irr galaxies, are only a very small fraction of sources, and thus are not represented in this analysis. On the other hand, more than 92\% of Sa-Scd are SF. The fact that the fraction of Q sources in later-type galaxies is low has also been reported in previous studies (e.g. \citealt{Goto2003b}; \citealt{Moran2006}; \citealt{Wolf2009}; \citealt{Masters2010}; \citealt{Fraser-McKelvie2016}, and references therein). Numerical simulations have shown that, in some cases, spiral galaxies, even many Gyr after a dramatic decrease of their star formation, are able to maintain their spiral arm structure \citep{Bekki2002}. In a volume limited sample of 5,433 spiral galaxies from the Galaxy Zoo (GZ1) clean catalogue \citep{Lintott2008}, selected from the SDSS Data Release 6, with redshift \mbox{0.03 < z < 0.085}, \citet{Masters2010} characterised 4-8\% of them as ``red'' spirals. These galaxies were found to be dominated by old stellar populations, while their SFR is lower compared to the main population of the same morphology bin. In a more recent study by \citet{Shimakawa2022}, investigating a sample of 1,100 spiral galaxies with \mbox{0.01 < z < 0.3}, from the Hyper Suprime-Cam Subaru Strategic Program \citep[HSC SSP;][]{Aihara2018}, found that 5\% are passive, having identical characteristics (in stellar populations) to the typical Q galaxies, despite the different morphologies.
These results are in quite good agreement with the corresponding Q late-type population (8\%) of the LTGs we find in the GAMA sample (see bottom-left panel of Fig.~\ref{fig:m-s/morph}). The SFR of these galaxies is found to be, in most cases, systematically lower than their SF counterparts of the same stellar mass bin. After visual inspection, as can be seen in Fig.~\ref{fig:giH}, these galaxies are spirals with a redder view and smoother spiral arms, in agreement with the findings by \citet{Shimakawa2022}. \citet{Goto2003b} refers to the red spirals as a population in transition between red E/S0 galaxies in low-redshift clusters and blue spirals frequent in higher-redshift clusters. 

The Q galaxy population is more numerous in earlier-type galaxies. However, although SF galaxies are, traditionally, considered to be mainly spirals and irregulars, we also observe a non-negligible population of early-spirals and ellipticals being SF too; a typical E with ongoing star formation is presented in Fig.~\ref{fig:giH}. About 43\% (45\% according to their sSFR) of the S0 galaxies are SF, while this fraction is 50\% (48\% according to sSFR) for Es. If we take into account the LBS galaxies as well, for which the vast majority are SF, we find a large population (68\%) of galaxies with elliptical or spheroidal morphology in the area of active star formation. Previous studies, such as \citet{Fukugita2004}, \citet{Schawinski2009}, \citet{Rowlands2012} have found EW(H$\alpha$) of ETGs that correspond to star-forming galaxies. \citet{Bitsakis2019} and \citet{CanoDiaz2019} find 18.5\% and 28\% of the ETGs being actively star-forming, respectively. Although these fractions are lower than the one found in our work, \citet{Bitsakis2019} although using a slightly more stringent criterion to define their quiescent sources (log sSFR < -11.5 yr$^{-1}$), their target selection criteria lead to a lack of lower mass galaxies ($10^8 - 10^{10}$ M$_\odot$). Similarly, \citet{CanoDiaz2019} uses a different criterion, a mix of the BPT \citep{BPT1981} and the WHAN diagram. In morphologically selected samples of E galaxies, \citet{Yi2005}, \citet{Kaviraj2007} and \citet{Schawinski2007} conclude that, at least, 15\%-30\% of the sources show evidence of recent star formation activity.

The high fraction of SF ETGs found in our work, might be affected by the exclusion, by the WHAN classification, of Q sources with weak emission lines, below the S/N=3 threshold.
What is noteworthy though, is the fact that the SF galaxies of any morphological bin show very similar SFMS. The corresponding scaling relations are discussed in the following paragraph.
   
\subsection{The place of SF early- and Q late-type galaxies in the local Universe}\label{place}

Previous studies (e.g. \citealt{Davies2019}; \citealt{Paspaliaris2021}; and references therein) have shown that the SFMS is, mainly, occupied by spiral galaxies, with the area below consisting, mainly, of E galaxies. Galaxies of intermediate Hubble-stages (i.e. S0, Sa) show a wider dispersion occupying both regimes as well as the area between them (often referred to as the green valley). The numerous SF ETGs found in the GAMA sample let us introduce a different perception of this conception.

   \begin{figure}[t!]
   \centering
   \includegraphics[width=0.48\textwidth]{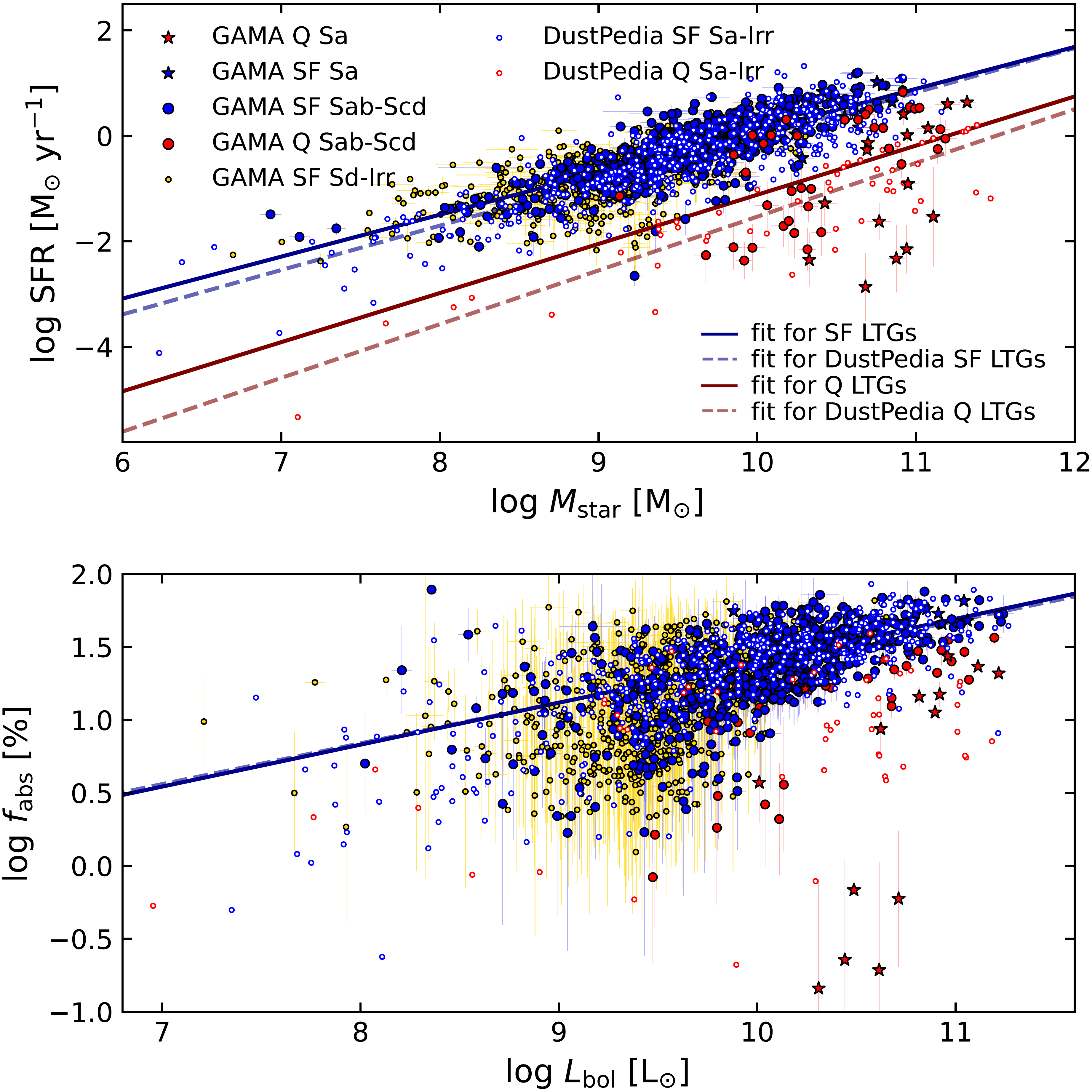}
   \caption{SFR vs $M_\text{star}$ (\textit{top panel})
   and $f_\text{abs}$ vs $L_\text{bol}$
   (\textit{bottom panel}) for the late-type (Sa-Scd and Sd-Irr) galaxies in the GAMA and DustPedia samples. SF Sa-Scd and Sd-Irr galaxies are represented with blue and yellow symbols, respectively, while Q Sa-Scd galaxies (given that Q Sd-Irr are under-represented and are not shown here; see Sect.~\ref{sec:classification}) are shown with red symbols. Blue and red stars stand for the GAMA SF and Q Sa galaxies, while Sab-Scd sources are represented by circles. Open symbols show the DustPedia Sa-Irr galaxies. All values are plotted along with their corresponding uncertainties. Solid lines are the linear fits for the GAMA SF and Q galaxies (blue and red, respectively), while the dashed lines are the linear fits for the DustPedia SF and Q galaxies (blue and red, respectively).
   }
   \label{fig:star_fabs_sp}
   \end{figure}

In Fig.~\ref{fig:star_fabs_sp} we plot the SFR as a function of the stellar mass (top panel), as well as the dust-to-bolometric luminosity ratio ($f_\text{abs}$ hereafter; discussed in detail in \citealt{Bianchi2018}) as a function of the bolometric luminosity (bottom panel) for LTGs. $f_\text{abs}$ is a quantity that provides an estimate of the amount of radiation that is reprocessed by dust. For comparison, representing the very local universe, the DustPedia galaxies are overplotted.
We note that the linear fits between the corresponding quantities of each subset presented in this paper are estimated using the python UltraNest\footnote{\url{https://johannesbuchner.github.io/UltraNest/}} package \citep{Buchner2021}. UltraNest derives the posterior probability distributions and the Bayesian evidence with the nested sampling Monte Carlo algorithm MLFriends \citep{Buchner2016,Buchner2019}. This method is a robust way to estimate the scaling relations of two properties taking into account the uncertainty of the data in both axes and being able to obtain the intrinsic scatter of the data and its uncertainties.

As expected, the vast majority of the Sa-Irr galaxies are distributed along the SFMS (top panel). The SFMS, as determined by the subset of GAMA SF Sa-Irrs, is described by the linear regression:
\begin{equation}
    \text{log}(\textit{SFR}[M_{\odot}\text{yr}^{-1}]) = 0.79^{+0.01}_{-0.01}~\text{log}(M_\text{star}[M_{\odot}]) - 7.83^{+0.13}_{-0.13}
\end{equation}
(blue solid line) and shows a very strong correlation (Pearson correlation coefficient $\rho$ = 0.8) and with an intrinsic scatter of 0.29$^{+0.01}_{-0.01}$ dex.
This is very similar to what is found for the local universe with the DustPedia relation being:
\begin{equation*}
    \text{log}(\textit{SFR}[M_{\odot}\text{yr}^{-1}]) = 0.84^{+0.02}_{-0.03}~\text{log}(M_\text{star}[M_{\odot}]) - 8.41^{+0.25}_{-0.21}
\end{equation*}
(dashed blue line; $\rho$ = 0.86) with the corresponding intrinsic scatter in this subsample being 0.38$^{+0.01}_{-0.01}$ dex. This also comes in agreement with other studies in the local universe.
For instance, \citet{Renzini2015} define the SFMS as a straight-line fit with a slope of 0.76$\pm$0.01 and an offset of -7.64$\pm$0.02 for $\sim$240,000 SDSS DR7 galaxies \citep{Abazajian2009}, lying at 0.02 < z < 0.085. Similarly, \citet{CanoDiaz2019} found a slope of 0.79$\pm$0.01 for the SF Sbc-Irr subset of their sample and 0.74$\pm$0.01 for their full sample of SF z$\sim$0 MaNGa galaxies. In a more recent study, \citet{Fraser-McKelvie2021}, for their \mbox{9.0 < log($M_\text{star}$[M$_{\odot}$]) < 10.0}, z $\sim$ 0 galaxies of the SAMI DR3 and MaNGA DR15 GALEX-Sloan-WISE legacy catalogue 2,  found a slope of 0.67. 

   \begin{figure}[t!]
   \centering
   \hspace{-0.9cm}\includegraphics[width=0.48\textwidth]{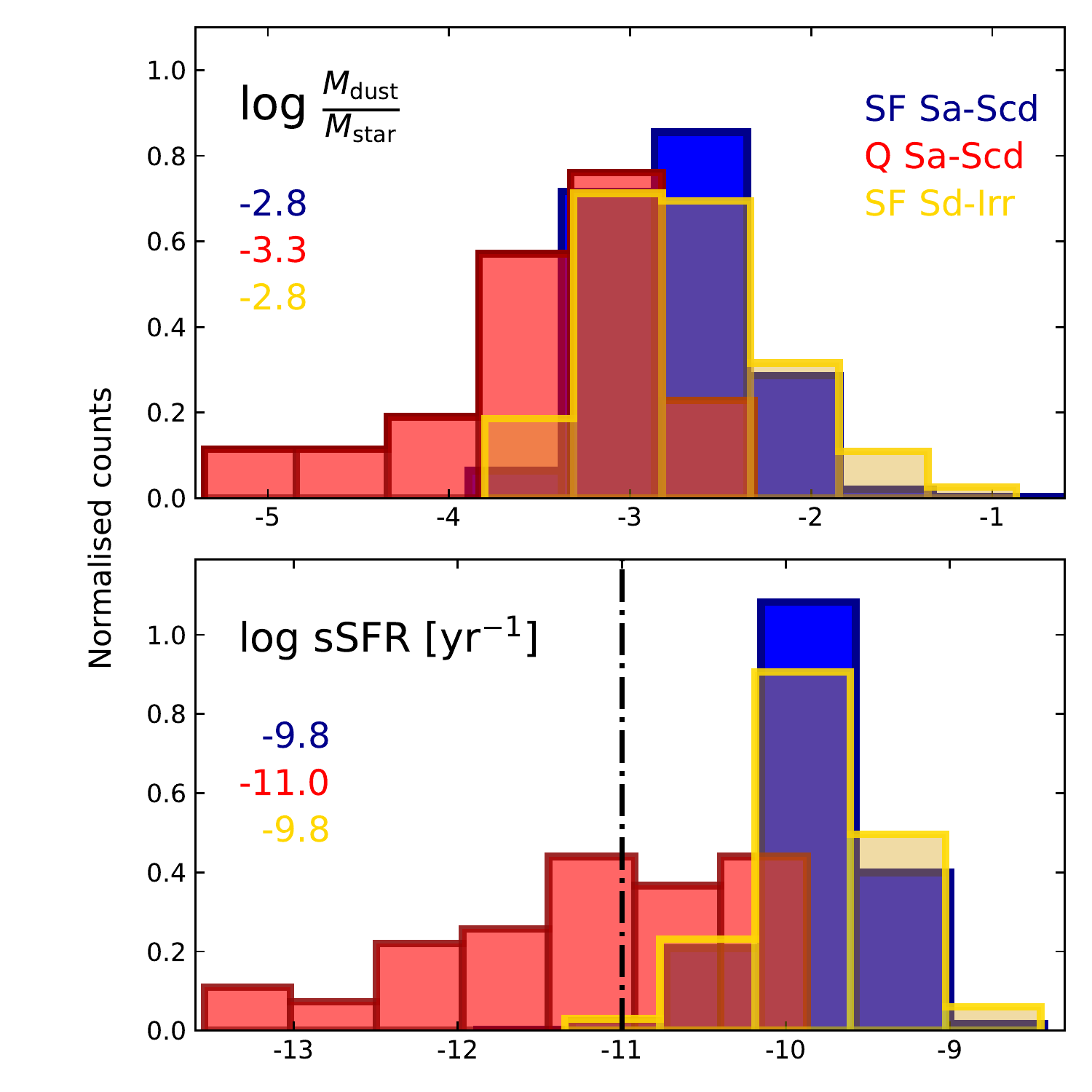}
   \caption{Comparison of the distributions of the \texttt{CIGALE}-derived dust-to-stellar-mass ratio  (\textit{top panel}) and sSFR (\textit{bottom panel}) for the different subclasses of late-type galaxies. The normalised distributions for the SF Sa-Scd, the Q Sa-Scd galaxies, and the SF Sd-Irr are shown in blue, red, and yellow, respectively. The median value of each parameter, for each population is given in the plot with the corresponding colour. The vertical dash-dotted line indicates the sSFR = 10$^{-11}$ yr$^{-1}$ threshold.}
   \label{fig:props_sp}
   \end{figure} 

Concerning the Q LTGs of the GAMA sample we find a weaker correlation ($\rho$ = 0.57) of: 
\begin{equation}
    \text{log}(\textit{SFR}[M_{\odot}\text{yr}^{-1}]) = 0.95^{+0.22}_{-0.20}~\text{log}(M_\text{star}[M_{\odot}]) - 10.59^{+2.30}_{-2.25}
\end{equation}
(red solid line), with a scatter of 0.82$^{+0.10}_{-0.08}$ dex. Considering the LTGs in the DustPedia sample with sSFR lower than 10$^{-11}$ yr$^{-1}$ as being Q systems (red open circles) we calculate a relation of:
\begin{equation*}
    \text{log}(\textit{SFR}[M_{\odot}\text{yr}^{-1}]) = 1.03^{+0.07}_{-0.06}~\text{log}(M_\text{star}[M_{\odot}]) - 11.88^{+0.73}_{-0.61}
\end{equation*}
(red dashed line) with a Pearson correlation coefficient of 0.88 and a scatter of 0.29$^{+0.01}_{-0.01}$ dex. Given that the scatter is quite large, the correlations are quite similar, though the GAMA sample shows a shift to higher SFR values. This is probably due to the fact that, using the WHAN diagram as the criterion to distinguish between SF and Q in the GAMA sample, we find some Q galaxies being above sSFR=10$^{-11}$ yr$^{-1}$, resulting in dragging the relation into higher SFR values.      
   
In the $f_\text{abs}$-$L_\text{bol}$ space (bottom panel) the distribution of the SF LTGs, in the GAMA sample, is described by the relation
\begin{equation}
    \text{log}(f_\text{abs}[\%]) = 0.28^{+0.01}_{-0.01}~\text{log}(L_\text{bol}[L_{\odot}]) - 1.35^{+0.13}_{-0.13}
\end{equation}
(blue solid line) with the intrinsic scatter being 0.14$^{+0.01}_{-0.01}$ dex and a correlation coefficient $\rho$ = 0.60.
The corresponding relation for the DustPedia sample is:
\begin{equation*}
    \text{log}(f_\text{abs}[\%]) = 0.29^{+0.01}_{-0.01}~\text{log}(L_\text{bol}[L_{\odot}]) - 1.45^{+0.14}_{-0.13},
\end{equation*}
(blue dashed line; $\rho$ = 0.74) which, given the scatter of the data (0.17$^{+0.01}_{-0.01}$ dex), is in excellent agreement with GAMA. The correlation coefficient ($\rho$ = 0.4) suggests that there is no correlation between $f_\text{abs}$ and $L_\text{bol}$ for the Q LTGs.

In order to further investigate the similarities and differences between Q and SF LTGs we examine their dust mass content (normalised to their stellar mass) as well as their sSFR. We do so in Fig.~\ref{fig:props_sp} with the dust-to-stellar mass ratio distributions presented in the top panel and the sSFR distributions in the bottom panel. From the top panel it is notable that SF Sa-Scd and SF Sd-Irr show similar distributions of dust-to-stellar mass ratio with the Q Sa-Scd showing significantly lower ratio. The median values of the distributions are $1.6\times10^{-3}$, $1.9\times10^{-3}$, and $5\times10^{-4}$ for the SF Sa-Scd, the SF Sd-Irr, and the Q Sa-Scd, respectively, indicating that the Q Sa-Scd are the most dust-poor types of all LTGs. A similar picture is seen in the bottom panel of Fig.~\ref{fig:props_sp} with SF Sa-Scd and SF Sd-Irr showing similar distributions in sSFR with that of Q Sa-Scd significantly deviating to lower values. Their median values are 1.58$^{-10}$ yr$^{-1}$ for SF Sa-Scd and SF Sd-Irr and 10$^{-11}$ yr$^{-1}$ for Q Sa-Scd. All the above lead to the conclusion that the Q Sa-Scd have less dust content, compared to their SF counterparts, but also they are, currently, forming new stars, at lower rates than the other types. Their stellar content will be revisited later in this paper (Sect.~\ref{sec:populations}) where the median SEDs will be presented.

The existence of a bar in galaxies has been suggested as a possible mechanism for galaxy quenching \citep{Kormendy&Kennicutt2004,Masters2011,Kruk2018,Fraser-McKelvie2020}. Additionally, several studies (e.g. \citealt{Masters2010}; \citealt{Fraser-McKelvie2018}; \citealt{Pak2019}) find a high fraction of passive spiral galaxies (up to $\sim$78\%) hosting a bar component, proposing that perturbation from the bar could cause star-formation quenching in these galaxies. In our sample we find that 12\% of the SF and 25\% of the Q Sa-Scd galaxies are barred. These results are in quite good agreement with the corresponding results found in \citet{Geron2021} showing that 13\% of their SF and 22\% of their Q galaxies possess a strong bar. Moreover, as in \citet{Geron2021}, we also find that barred spiral galaxies (i.e. SBa-SBcds) have enhanced SFR compared to their unbarred analogues (0.88 M$_{\odot}\text{yr}^{-1}$ and 0.63 M$_{\odot}\text{yr}^{-1}$, respectively), indicating probable rapid evolution of barred galaxies with the bar expediting the ceasing procedure.
\citet{George2019} found that 54.3\% of barred galaxies in the local Universe are quenched, with this fraction becoming even larger (66.5\%) for galaxies with $M_\text{star}$ > 10$^{10.2}$ M$_{\odot}$. In our sample 13.2\% of barred galaxies are quiescent with this fraction becoming 74.3\% for galaxies with $M_\text{star}$ > 10$^{10.2}$ M$_{\odot}$. The enhanced bar fraction in Q spirals could be a possible indication that bars (or their creation mechanisms) are, at some level, responsible for the process of the ceasing of star formation. \citet{CombesSanders1981} suggested that bars are able to redistribute the material of disc galaxies by funnelling gas towards the galaxy centre. Such a process could have occurred in Q spirals in our sample, removing the available cold gas reservoir from their disc and making it available in the central regions for the induction of starburst/AGN activity followed by quenching (e.g. \citealt{Knapen2002}; \citealt{Jogee2005}).

\begin{figure}[t!]
   \centering
   \includegraphics[width=0.48\textwidth]{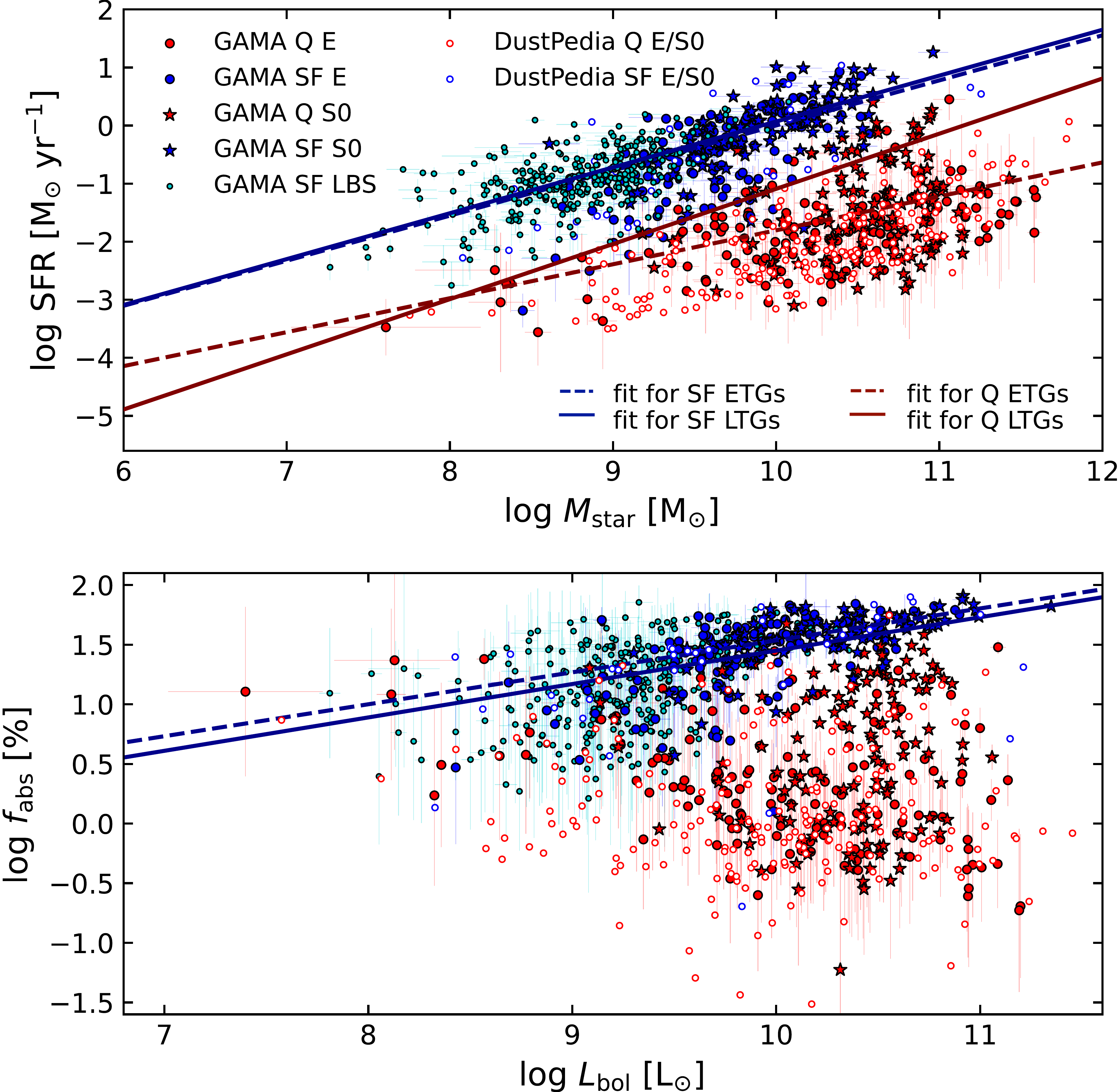}
   \caption{SFR vs $M_\text{star}$ (\textit{top panel})
   and $f_\text{abs}$ vs $L_\text{bol}$
   (\textit{bottom panel}) for the early-type (E/S0 and LBS) galaxies in the GAMA and DustPedia samples. SF E/S0 and SF LBS galaxies are represented with blue and cyan symbols, respectively, while Q E/S0 galaxies (given that Q LBS are under-represented and are not shown here; see Sect.~\ref{sec:classification}) are shown with red symbols. Blue and red stars stand for the GAMA SF and Q S0 galaxies, while E sources are represented by circles. Open symbols show the DustPedia E/S0 galaxies. All values are plotted along with their corresponding uncertainties. Solid lines are the linear fits for the GAMA SF and Q galaxies (blue and red, respectively), while the dashed lines are the linear fits for the DustPedia SF and Q galaxies (blue and red, respectively).
   }
   \label{fig:star_fabs_e}
   \end{figure}

As already presented in the top panels of Fig.~\ref{fig:m-s/morph}, local ETGs (LBS; left panel, and E/S0; right panel) occupy different loci in the SFR-$M_\text{star}$ plane, depending on their stellar mass and their star-forming activity. To further investigate similarities and differences between these ETG populations (SF E/S0, Q E/S0 and SF LBS), in Fig.~\ref{fig:star_fabs_e} we plot, as for the case of LTGs, the SFMS (top panel), as well as the $f_\text{abs}$ as a function of the bolometric luminosity (bottom panel). The fact that SF LBS galaxies follow the same distribution with the SF Es confirms the argument of \citet{Kormendy2009} that the population of high surface brightness concentrated galaxies (such as LBSs) is similar to the giant elliptical population. The best linear fits for the two different populations (SF and Q) are plotted in the top panel of Fig.~\ref{fig:star_fabs_e}, with the majority of the SF E/S0s along with the SF LBS galaxies constituting a separate distribution, following, very closely, a relation similar to the SFMS of the local SF spiral galaxies [Eq. (1); blue solid line]. The SFMS of the GAMA SF ETGs ($\rho$ = 0.73; blue dashed line) is described by the following linear relation,
\begin{equation*}
    \text{log}(\textit{SFR}[M_{\odot}\text{yr}^{-1}]) = 0.78^{+0.03}_{-0.02}~\text{log}(M_\text{star}[M_{\odot}]) - 7.79^{+0.25}_{-0.23}
\end{equation*}
and their intrinsic scatter is found to be 0.36$^{+0.01}_{-0.01}$ dex.
This is in agreement with the findings by \citet{Wu2021} who have found a slope of 0.74 $\pm$ 0.01 and an offset of -7.22 $\pm$ 0.08 for their sample of SDSS DR7 SF ETGs.

   \begin{figure}[t!]
   \centering
   \hspace{-0.9cm}\includegraphics[width=0.47\textwidth]{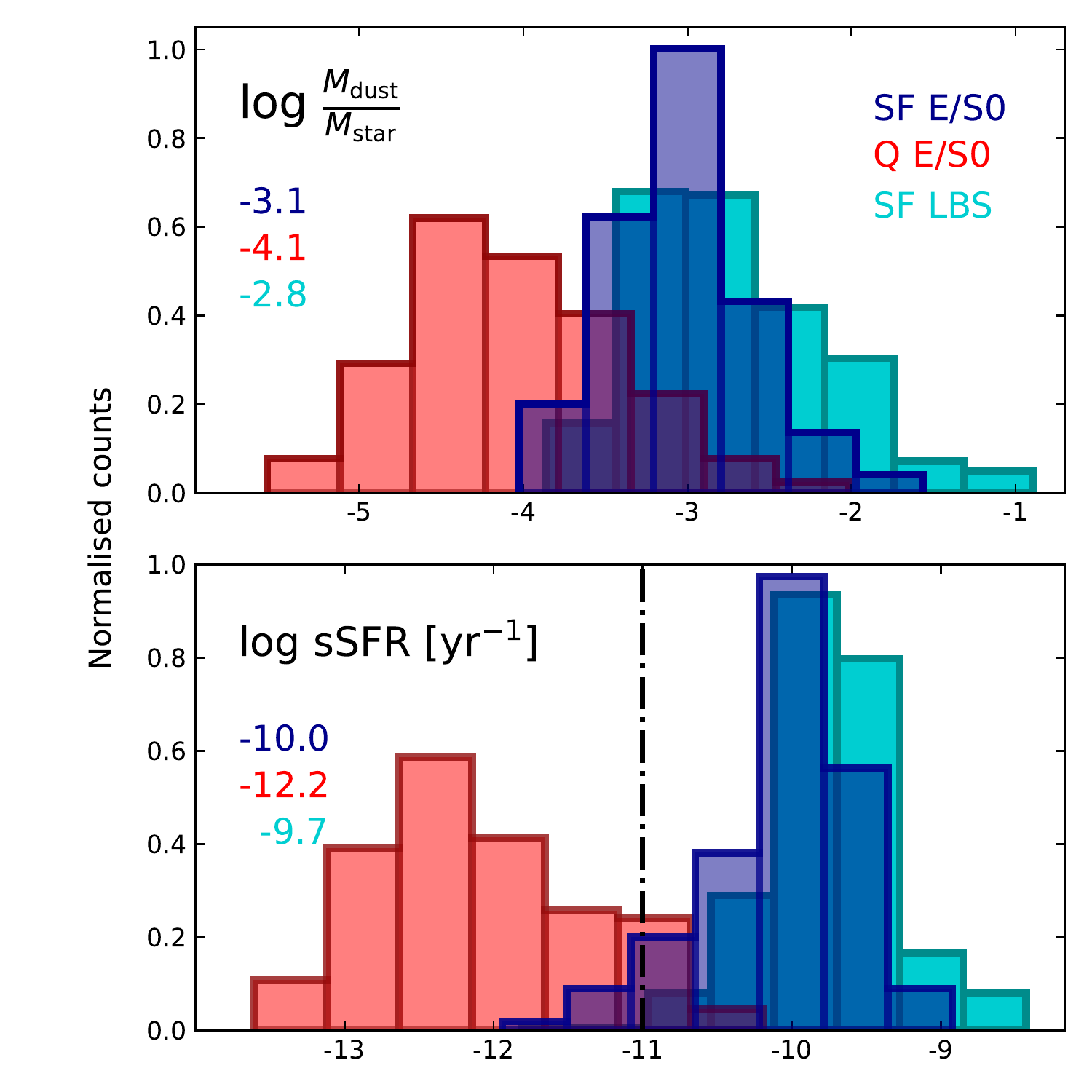}
   \caption{Comparison of the distributions of the \texttt{CIGALE}-derived dust-to-stellar-mass ratio  (\textit{top panel}) and sSFR (\textit{bottom panel}) for the different subclasses of early-type galaxies. The normalised distributions for the SF E/S0, the Q E/S0 galaxies, and the SF LBS are shown in blue, red, and cyan, respectively. The median value of each parameter, for each population is given in the plot with the corresponding colour. The vertical dash-dotted line indicates the sSFR = 10$^{-11}$ yr$^{-1}$ threshold.}
   \label{fig:props_e}
   \end{figure}

   \begin{figure*}[t!]
   \centering
   \includegraphics[width=\textwidth]{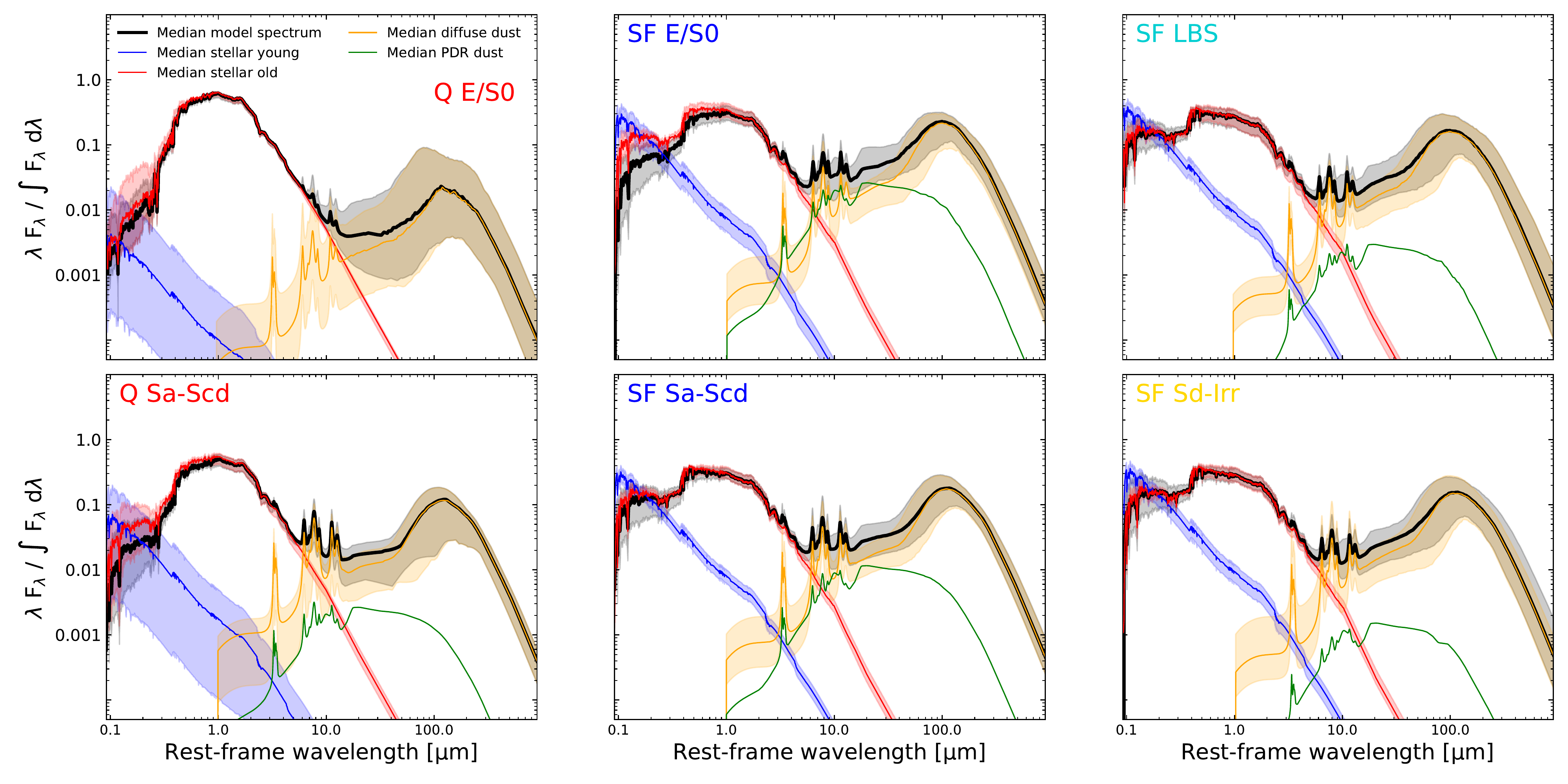}
   \caption{Median SEDs of the GAMA ETGs (\textit{top panels}) and the LTGs (\textit{bottom panels}). The median total luminosity of each sub-class is indicated in black, while the old and young stellar components are shown in red and blue, respectively. The orange curve stands for the diffuse dust, while dust in PDRs is shown in green. Shaded areas correspond to the 16th-84th percentile range.
   }
   \label{fig:seds}
   \end{figure*}

A second correlation is present in this plot for the Q E/S0s, with a larger scatter though (0.64$^{+0.04}_{-0.03}$ dex) and lower Pearson correlation coefficient $\rho$ = 0.58. GAMA Q E/S0s occupy the same locus with the DustPedia Q E/S0s in the SFR-$M_\text{star}$ plane. The corresponding linear regression to the GAMA Q E/S0s gives: 
\begin{equation*}
    \text{log}(\textit{SFR}[M_{\odot}\text{yr}^{-1}]) = 0.59^{+0.08}_{-0.08}~\text{log}(M_\text{star}[M_{\odot}]) - 7.65^{+0.84}_{-0.89},
\end{equation*}
(red dashed line). Compared to the Q LTGs [Eq. (2); red solid line] we see that the Q ETGs occupy a slightly different space in the SFR-$M_\text{star}$ plane with lower values of SFR per stellar mass bin.

\citet{Nersesian2019b} have shown that $f_\text{abs}$ is larger in intermediate spiral galaxies (Sb-Sc), where $\sim35\%$ of their intrinsic luminosity is affected by the dust. On the other hand, it is also shown in that study that, pure elliptical galaxies exhibit very low $f_\text{abs}$ values ($\sim2\%$). \citet{Bianchi2018} explored the dependence of this quantity on the bolometric luminosity, finding a positive correlation for the late spirals. The Es though were found to have no correlation. Similarly, in the bottom-panel of Fig.~\ref{fig:star_fabs_e} we plot $f_\text{abs}$ as a function of the bolometric luminosity with the nomenclature being the same as in the top panel of the same figure. As was also observed in the SFR-$M_\text{star}$ plane, again we find the SF E/S0s, along with the LBS galaxies, constituting a separate distribution in the $f_\text{abs}$-$L_\text{bol}$ plane. Regardless their elliptical morphology they follow a positive trend (blue dashed line; $\rho$ = 0.62), described by the relation:
\begin{equation*}
    \text{log}(f_\text{abs}[\%]) = 0.27^{+0.02}_{-0.02}~\text{log}(L_\text{bol}[L_{\odot}]) - 1.12^{+0.17}_{-0.18}
\end{equation*}
with intrinsic scatter 0.12$^{+0.01}_{-0.01}$.
This correlation is in great agreement with what is found for the SF LTGs (Eq. (3); solid blue line).
No clear correlation seems to be present for the $f_\text{abs}$ with the $L_\text{bol}$ for the Q E/S0 galaxies ($\rho$ = -0.12). 

To further investigate the differences in the physical properties between the SF and Q ETGs, in Fig.~\ref{fig:props_e} we present the histograms of the dust-to-stellar-mass ratio (top panel) and the sSFR (bottom panel). The histogram of the dust-to-stellar-mass ratio suggests that among the three populations (SF LBS, Q E/S0, and SF E/S0) it is the SF LBS that show the highest relative dust content, compared to their stellar mass, with a median value of the dust-to-stellar-mass ratio of 0.0016. The corresponding ratio of the SF E/S0s is $8\times10^{-4}$, whilst the Q E/S0 galaxies are the most dust deficient for their stellar mass (a ratio of $8\times10^{-5}$).
The median ratio we find for the SF ETGs (0.0016) is in excellent agreement with the findings by \citet{Rowlands2012} who computed the ratio for a Herschel-Astrophysical Terahertz Large Area Survey (H-ATLAS; \citealt{Eales2010})/GAMA matched sample, highly contaminated by UV/optical blue ETGs. In another volume-limited sample of 62 ETGs of the Herschel Reference Survey (HRS), with 24\% of them having detected dust emission, \citet{Smith2012} found a 
mean ratio of $5\times10^{-5}$. 
This result is in agreement (within 0.2 dex difference) with the ratio we find for the Q E/S0s. 
   
By examining the distributions of the sSFR for the different earlier-type galaxy populations, we find the SF E/S0s having a median sSFR of $10^{-10.0}$ yr$^{-1}$ compared to the Q ones ($6\times10^{-13}$ yr$^{-1}$). As was indicated by their high dust-to-stellar-mass ratio SF LBSs is found to be the most actively star-forming early-type population with a median value of $2\times10^{-10}$ yr$^{-1}$. The mean sSFR averaged over the last 100 Myr for blue ETGs reported by \citet{Rowlands2012} is $1.2\times10^{-11}$ yr$^{-1}$ (1 dex lower from the mean sSFR of SF E/S0s). Respectively, the median sSFR for the DustPedia E/S0s (which in vast majority are Q) is found to be $6.3\times10^{-13}$ yr$^{-1}$, the same for GAMA Q E/S0s.

As was also reported in previous studies for blue ETGs (e.g. \citealt{FerrerasSilk2000}; \citealt{Kaviraj2007}; \citealt{Sampaio2022}) SF ETGs are mainly found having lower masses compared to their Q counterparts. In GAMAnear we found 90\% of SF ETGs lying in the $1.6\times10^{8}$ M$_{\odot}$ to $2.0\times10^{10}$ M$_{\odot}$ range of $M_\text{star}$, while Q ETGs exceed to stellar masses up to $4.0\times10^{11}$ M$_{\odot}$. A similar behaviour is observed for spirals, where Q Sa-Scds are mainly found at the high-mass end. Furthermore, as we move from LTGs to ETGs we find the dust-to-stellar-mass ratio to decrease for SF galaxies (0.3 dex) and more significantly for Q galaxies (0.8 dex). This decreasing trend is in agreement with the findings by \citet{Cortese2012} studying the dust scaling relations of a HRS volume- and magnitude-limited sample of $\sim$300 galaxies.

It is worth noticing that the number of Q Sa-Scds with $f_\text{abs}$ larger than 10\% ($\sim$70\%) is significantly larger than the corresponding fraction for the Q E/S0s ($\sim$21\%). This suggests that the dust effects in Q Sa-Scds are more important than in Q E/S0s. This is to be expected since Q Sa-Scds are found to be more dusty for their stellar masses, with their corresponding ratio being 0.8 dex higher than the one for Q E/S0s (see Figs~\ref{fig:props_sp} and \ref{fig:props_e}).

\section{Stellar populations in SF early- and Q late-type galaxies and their role in dust heating}\label{sec:populations}

Investigating the stellar populations in galaxies and the way their radiation interacts with the dust grains is essential to understanding the energy balance that is taking place inside galaxies. The approach that we utilise, by modelling multi-wavelength SEDs of galaxies with \texttt{CIGALE}, allows us to parametrise the stellar content of galaxies into two broad categories, namely, an old and a young stellar population. The old stars are modelled with an exponentially declining SFR with the e-folding time (0.5-20 Gyr) and the age (2-12 Gyr) as free parameters. The young stars are formed after a bursting or quenching event of star formation, started 200 Myrs ago, with the ratio of the SFR before and after the event being a free parameter. The choice of a constant moment for the latter event was indicated by \citet{Ciesla2016}  who found that the shape of the SED is not sensitive to variations of this parameter (see also \citealt{Nersesian2019b}). 

   \begin{figure*}[t!]
   \centering
   \includegraphics[width=\textwidth]{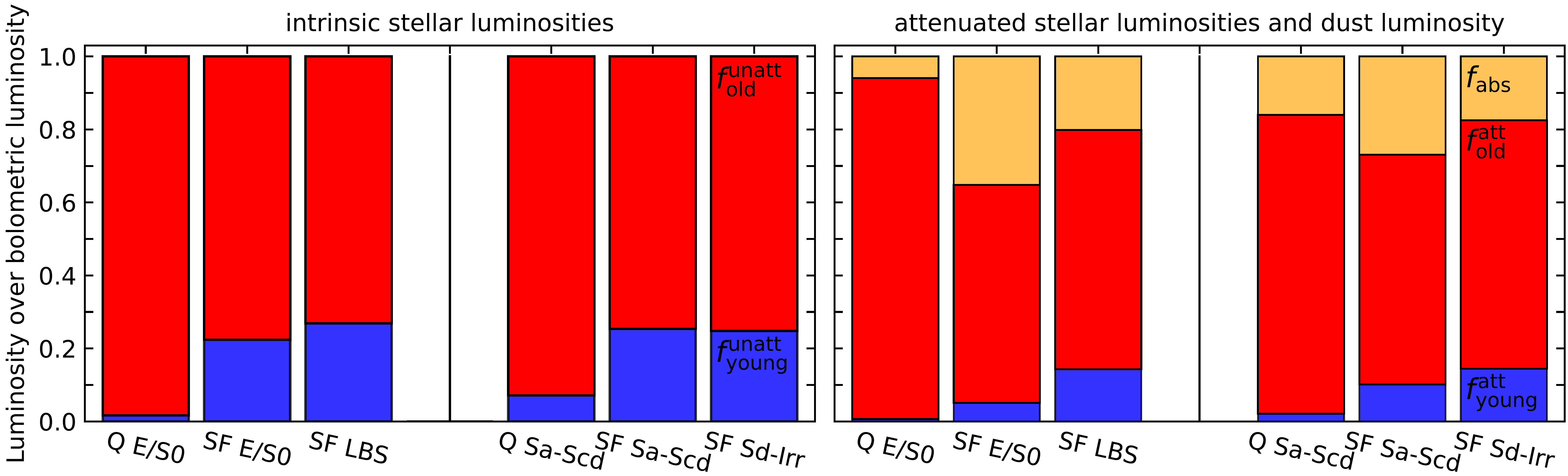}
   \caption{The fraction of the old (red) and young (blue) stellar populations to the mean unattenuated ($f_\text{old,young}^\text{unatt}$ = $L_\text{old,young}^\text{unatt}$/$L_\text{bol}$; \textit{left panel}) and to the mean attenuated luminosity ($f_\text{old,young}^\text{att}$ = $L_\text{old,young}^\text{att}$/$L_\text{bol}$; \textit{right panel}), to the bolometric luminosity, per galaxy subset. In the \textit{right panel} the ratio of the mean dust luminosity over the bolometric luminosity is also shown in orange colour ($f_\text{abs}$ = $L_\text{dust}$/$L_\text{bol}$). 
   }
   \label{fig:bars}
   \end{figure*}

\citet{Nersesian2019b} studied the fraction of the luminosities of the two stellar components to the total luminosity, as well as their effect in the dust heating (the fractions of the stellar luminosities absorbed by dust) for the galaxies in the DustPedia sample, as a function of their morphology. This analysis suggested that the old stars in local galaxies, of all morphological types, always constitute the dominant population in terms of luminosity, with the contribution of young stars being less significant following an evolutionary sequence, decreasing from later- to earlier-types of galaxies. On average, it was found that, the ratio of the luminosity of the young stars to the bolometric luminosity of the galaxy is 25\% for Sb-Irr morphologies, while it drops to less than 10\% for E/S0 morphologies. Despite this trend where ETGs, in the local universe, are mostly composed of older stars and with LTGs hosting larger fractions of young stars, other studies have reported the existence of galaxies where populations, significantly, depart from this general picture. 
\citet{Schawinski2009}, investigating a sample of visually identified blue ETGs, selected from Galaxy Zoo and SDSS DR6, found that a young stellar component is present in this galaxy population. Measuring the Balmer absorption-line index H$\delta_\text{A}$ (tracer of the recent star formation) and the break at $\sim$4000 ${\AA}$ (tracer of mean stellar age) \citet{Masters2010} found the range of stellar ages in passive spirals being similar to typical ETGs and older than the one of star-forming spirals. Similarly, \citet{Rowlands2012}, estimating the r-band light-weighted age of passive spiral galaxies found that they are mostly dominated by old stars with current star-formation activity well below the normal spirals. 

Taking advantage of the fact that the galaxies in our sample are categorised in SF and Q in an unbiased and independent way (optical spectra; see Fig.~\ref{fig:whan}) we are able to examine how the relative contributions of the two stellar populations (old and young) change, not only with morphology, but also, with star-formation activity. 
In Fig.~\ref{fig:seds} we present median SEDs of ETGs (Q E/S0, SF E/S0, SF LBS; top panels, left to right) and LTGs (Q Sa-Scd, SF Sa-Scd, SF Sd-Irr; bottom panels, left to right). In the panels describing the different types, apart from the total median SED, the various components comprising the SED (old and young stellar population, diffuse and PDR dust emission) are also shown. A visual inspection of the SEDs reveals a dependence of their shape with, both, the morphology and star-formation activity. 

Amongst the ETGs (top row in Fig.~\ref{fig:seds}) the median SEDs of SF E/S0s and SF LBSs are very similar, while that of Q E/S0s deviates a lot. The similarity of the first two types of SEDs is more evident when comparing the flux at 100 $\mu$m (a difference of 0.14 dex) and at 1.0 $\mu$m (a difference of 0.06 dex), an indication of similar dust and stellar emission, respectively. One needs to notice here that LBSs are more compact and less massive systems, compared to SF E/S0s, so, even though the shape of their SED (normalised by their bolometric luminosity) is similar, they are expected, on average, to be less luminous. The similarities in the SEDs suggest that SF LBSs and SF E/S0s (especially Es which have similar overall morphology with LBSs) belong to the same population of galaxies with SF LBSs occupying the low-mass end of this galaxy population. Q E/S0s, on the other hand (top-left panel), show a very different SED with the FIR emission being weaker, compared to the other two types, and with the stellar SED dominated by emission from old stars. The contribution from young stars in Q E/S0s is almost negligible. Q ETGs, for instance, compared to SF ETGs, have 0.3 dex higher flux at 1.0 $\mu$m and 1.1 dex lower flux at 100 $\mu$m. The lack of young stars in Q E/S0s is also reflected by the absence of PDR emission (linked to emission originating from star-forming sites) in the IR part of the SED, compared to the other two types where this type of emission has a significant contribution, especially in the case of SF E/S0s.

Although spiral galaxies (Q/SF Sa-Scd) show a very similar FIR emission (left and middle bottom panels) with a difference of only 0.21 dex at 100 $\mu$m, they differ a lot in their optical and FUV emission (e.g. 0.75 dex difference at $\sim$0.1 $\mu$m). Similar to the case of SF E/S0s and SF LBSs which are found to have very similar median SEDs, in the LTGs case, SF Sd-Irrs have almost identical median SED with the SF Sa-Scds (0.03 dex and 0.06 dex difference at 1.0 and 100 $\mu$m, respectively). This suggests that (as opposed to ETGs where one can see differences between Q and SF in both the stellar and the dust emission) LTGs galaxies differ mostly in their stellar content with SF types hosting significantly more young stars.

All the aforementioned properties are better quantified in Fig.~\ref{fig:bars}, where we present the relative contributions of each stellar component (old and young; red and blue colours, respectively) to the total bolometric luminosity. In this figure, the leftmost panels show the relative contributions, to the bolometric luminosity, of the pure, unattenuated, stellar components (i.e. if there was no dust present in the galaxies) while in the rightmost panels the effects of dust are taken into account. In both panels, the relevant information for both ETGs (Q E/S0, SF E/S0, SF LBS) and LTGs (Q Sa-Scd, SF Sa-Scd, SF Sd-Irr) are presented. All the luminosity fractions are presented in Table~\ref{tab:old_young_dust}.

The first thing to notice from this plot is that, regardless of the morphological type or the star-forming activity, the old, more evolved, stellar population dominates the bolometric luminosity. This was also evident in \cite{Nersesian2019b} with more than $\sim$75\% of the bolometric luminosity, in all types of galaxies, originating from old stars, and with Es being the most extreme cases with this fraction being as high as 98\%. On the other hand, in Sbc to Irr galaxies the fraction of the luminosity coming from the young stars is more significant, ranging between 20\% and 30\% depending on the galaxy type. With the current analysis we can update these results, not only as a function of morphology, but also, as a function of the star-forming activity in galaxies. Concerning the ETGs we find that Q E/S0s have a negligible fraction of young stars (1\%), while SF E/S0s and SF LBSs show a quite substantial contribution of young stars (23\% and 25\%, respectively) comparable to what is seen for the most actively star-forming spiral galaxies). Nevertheless as was reported by \citet{Huang2009}, although the young stellar population is enhanced in SF Es (and SF S0s in our sample), they should not be considered as young objects, since their main stellar population is as old as the ones in the Q Es. LTGs, on the other hand, show a small fraction (4\%) of young stars when being quiescent (Q Sa-Scds) which gets higher (25\%) in their star-forming counterparts (SF Sa-Scds/Sd-Irrs).
\citet{Huang2009} found that due to their star-formation activity, SF Es have higher levels of dust attenuation. On the other hand, as was also discussed in paragraph \ref{place}, Q E/S0s are the most dust poor for their stellar mass, among the ETGs. A comparison of the dust content for the different types of LTGs is also provided in the same paragraph, finding that Q Sa-Scds are the most dust deficient. 

   \begin{table*}[t]
   \caption{Average values of the ratios of various combinations of the stellar and dust luminosity components extracted using the \texttt{CIGALE} SED fitting tool. The different ratios (also presented in \citealt{Nersesian2019b} and \citealt{Paspaliaris2021}) are defined as
   $f^\text{unatt}_\text{old} = L^\text{unatt}_\text{old}/L_\text{bol}$, $f^\text{unatt}_\text{young} = L^\text{unatt}_\text{young}/L_\text{bol}$, $f^\text{att}_\text{old} = L^\text{att}_\text{old}/L_\text{bol}$, $f^\text{att}_\text{young} = L^\text{att}_\text{young}/L_\text{bol}$,
   $f_\text{abs} = L_\text{dust}/L_\text{bol}$,
   $F^\text{att}_\text{old} = L^\text{att}_\text{old}/L^\text{unatt}_\text{old}$,
   $F^\text{abs}_\text{old} = L^\text{abs}_\text{old}/L^\text{unatt}_\text{old}$,
   $F^\text{att}_\text{young} = L^\text{att}_\text{young}/L^\text{unatt}_\text{young}$, $F^\text{abs}_\text{young} = L^\text{abs}_\text{young}/L^\text{unatt}_\text{young}$, $S^\text{abs}_\text{old} = L^\text{att}_\text{old}/L_\text{dust}$, and $S^\text{abs}_\text{young} = L^\text{att}_\text{young}/L_\text{dust}$, 
   where, $L^\text{unatt}_\text{old}$ and $L^\text{unatt}_\text{young}$
   are the unattenuated luminosities of the old and the young stars, $L_\text{bol}$ is the bolometric luminosity of each system ($L_\text{bol} = L^\text{unatt}_\text{old} + L^\text{unatt}_\text{young}$), $L_\mathrm{dust}$ is the dust luminosity, $L_\mathrm{old}^\mathrm{att}$ and $L_\mathrm{young}^\mathrm{att}$ are the attenuated luminosity of the old and young stars, and $L_\mathrm{old}^\mathrm{abs}$ and $L_\mathrm{young}^\mathrm{abs}$ are the luminosity of the old and young stars absorbed by dust. These ratios are shown for each galaxy subset.}
   \begin{center}
   \scalebox{0.94}{
   \begin{tabular}{l c c c c c c c c c c c}
\hline 
\hline 
   Galaxy Type &
   $f_\mathrm{old}^\mathrm{unatt}$ & $f_\mathrm{young}^\mathrm{unatt}$ & $f_\mathrm{old}^\mathrm{att}$ & $f_\mathrm{young}^\mathrm{att}$ & $f_\mathrm{abs}$ & $F_\mathrm{old}^\mathrm{att}$ & $F_\mathrm{old}^\mathrm{abs}$ &  $F_\mathrm{young}^\mathrm{att}$ & $F_\mathrm{young}^\mathrm{abs}$ & $S_\mathrm{old}^\mathrm{abs}$ & $S_\mathrm{young}^\mathrm{abs}$ \\
\hline
   Q E/S0 & 0.99 & 0.01 & 0.93 & 0.01 & 0.06 & 0.94 & 0.06 & 0.64 & 0.36 & 0.91 & 0.09 \\
   Q Sa-Scd & 0.96 & 0.04 & 0.82 & 0.02 & 0.16 & 0.88 & 0.12 & 0.40 & 0.60 & 0.76 & 0.24 \\
   SF LBS & 0.75 & 0.25 & 0.66 & 0.20 & 0.19 & 0.89 & 0.11 & 0.56 & 0.44 & 0.39 & 0.61 \\
   SF E/S0 & 0.77 & 0.23 & 0.50 & 0.05 & 0.35 & 0.76 & 0.24 & 0.26 & 0.74 & 0.53 & 0.47 \\
   SF Sa-Scd & 0.75 & 0.25 & 0.63 & 0.10 & 0.27 & 0.84 & 0.16 & 0.41 & 0.59 & 0.43 & 0.57 \\
   SF Sd-Irr & 0.75 & 0.25 & 0.68 & 0.14 & 0.18 & 0.90 & 0.10 & 0.58 & 0.42 & 0.41 & 0.59 \\
\hline
   \end{tabular}}
   \label{tab:old_young_dust}
   \end{center}
   \end{table*}

The interplay between the two stellar populations and dust (in the different galaxy types) is shown in the right panel of Fig.~\ref{fig:bars}. A decrease in the luminosity of both stellar populations is observed and this energy is absorbed and then re-emitted by dust (yellow bars). What is worth noticing from this plot is that although, Q E/S0s have low contribution to the total luminosity by dust, as expected, the contribution is very significant in SF E/S0s and SF LBSs, 35\% and 19\%, respectively.

When dust is considered the observed stellar luminosities are now suppressed as a result of the absorption of the stellar radiation by dust grains. The quantitative picture of this effect is described in the rightmost panels of Fig.~\ref{fig:bars} with blue and red bars indicating the mean ratios of the observed to the bolometric luminosities of the young and old stellar populations, respectively. Additionally, yellow bars indicate the fraction of the stellar radiation that is absorbed and gone into the dust heating ($f_\text{abs}$ = $L_\text{dust}$/$L_\text{bol}$). As also discussed in \citet{Bianchi2018} and \citet{Nersesian2019b} this quantity indicates the significance of the dust in galaxies and the
effectiveness of the dust grains in absorbing the stellar radiation, a combination of the total amount of dust, the geometry, and the strength of the ISRF (see also \citealt{Paspaliaris2021}). It is interesting to notice here that SF E/S0s show the highest fraction (35\%) of the stellar radiation absorbed by dust, followed by SF spirals (27\%), while SF LBSs and SF Sd-Irr have a similar fraction (19\% and 18\%, respectively). Q E/S0s have the lowest fraction $f_\text{abs}$ (6\%).

   \begin{figure}[t!]
   \centering
   \includegraphics[width=0.5\textwidth]{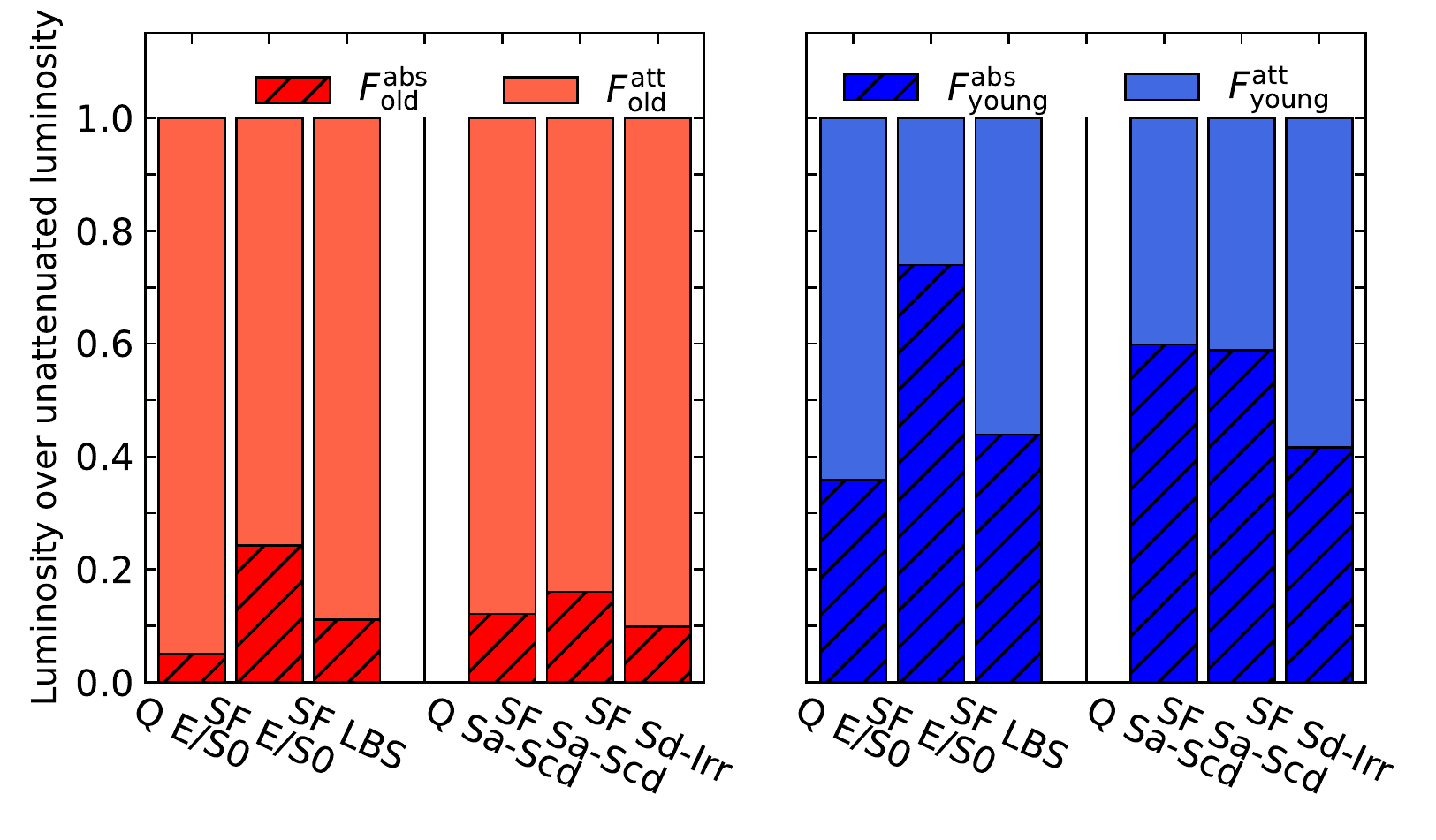}
   \caption{Mean values of the fraction of the luminosity of each stellar population (red for old and blue for young stars) used for the heating of the dust. The ratio of the dust-absorbed luminosity to the unattenuated luminosity of the corresponding stellar component is indicated by the dashed bars ($F_\text{old,young}^\text{abs}$ = $L_\text{old,young}^\text{abs}$/$L_\text{old,young}^\text{unatt}$). Solid bars represent the ratios of the attenuated luminosity to the unattenuated luminosity ($F_\text{old,young}^\text{att}$ = $L_\text{old,young}^\text{att}$/$L_\text{old,young}^\text{unatt}$). 
   }
   \label{fig:barsFabs}
   \end{figure}

The ratios of the stellar luminosity absorbed by the dust ($L_\text{old,young}^\text{abs}$) to the total, unattenuated, stellar luminosity ($L_\text{old,young}^\text{unatt}$) for each stellar component (old and young; $F_\text{old,young}^\text{abs}$) can provide an estimation of how effectively each stellar population can heat up the dust. These luminosity ratios are presented in Fig.~\ref{fig:barsFabs},with the left panel showing the relative contribution of the old stars and the right panel the relative contribution of the young stars to the dust heating, respectively, for the various types of ETGs and LTGs. The dashed part of each bar corresponds to the part of the intrinsic luminosity, of each stellar component, that is absorbed by the dust, while the solid part of each bar indicates the remaining, emitted directly by the stars, luminosity. 
At a first glance, it is obvious that, in all cases, young stars are the ones that donate a higher fraction of their energy to the heating of the dust (already noted in \citealt{Nersesian2019b} and \citealt{Paspaliaris2021}). 
Amongst the ETGs (Q E/S0, SF E/S0, and SF LBS) it is the SF E/S0 in which the stars are more efficiently heating the dust with old stars donating up to 24\% (and young stars up to 74\%) of their luminosity to heat up the dust. Between the two classes of LTGs (Q and SF) there is not much variation with the old stars donating $\sim$ 10\%-16\% while, surprisingly enough, the Q spirals give similar portion of their young-stellar luminosity to heat up the dust (60\% compared to 59\% for the SF spirals). The relative distribution of stars and dust inside galaxies is expected to affect the fraction of the radiation of each stellar population offered for the dust heating. The fact that the offered (absorbed) fraction of each stellar population is comparable in Q and SF spiral galaxies (which geometry is obviously very similar), while the corresponding percentages in Q and SF E/S0 galaxies are significantly different, may indicate differences in the internal structure of these sources (especially in SF Es), despite their similar overall structural characteristics.

   \begin{figure}[t!]
   \centering
   \includegraphics[width=0.485\textwidth]{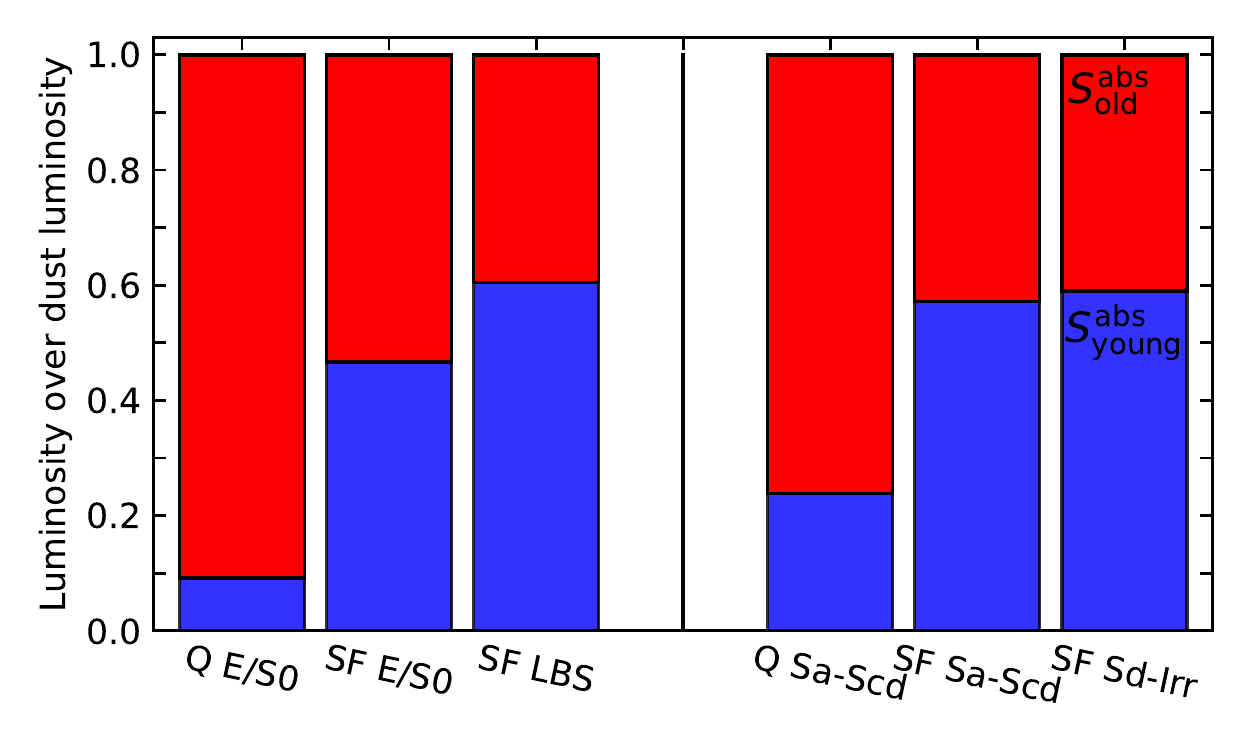}
   \caption{Mean values of the ratios of the old (red) and the young (blue) stellar luminosity absorbed by dust to the dust luminosity \mbox{($S_\text{old,young}^\text{abs}$ = $L_\text{old,young}^\text{abs}$/$L_\text{dust}$)}. 
   }
   \label{fig:bars_Sabs}
   \end{figure}

The relative contribution of each stellar population to the heating of the dust ($S_\text{old,young}^\text{abs}$) is shown in Fig.~\ref{fig:bars_Sabs}, with the red and blue bars representing the mean ratio of the absorbed, by the dust, luminosity originating from the old and the young stars ($L_\text{old,young}^\text{abs}$), respectively, to the total luminosity emitted by the dust ($L_\text{dust}$). It is interesting to notice that it is only the SF Sa-Scd, SF Sd-Irr galaxies and the SF LBSs in which the contribution of the young stars is, on average, the dominant source of dust heating with values of $S_\text{young}^\text{abs}$ of 57\%, 59\% and 61\%, respectively. In Q spirals and SF E/S0s there is a significant contribution in the dust heating from the young stars (24\% and 47\%, respectively), while the dust in Q E/S0s is heated, almost exclusively, by old stars ($S_\text{old}^\text{abs}$ of 91\%). All the relevant average fractions are presented in Table~\ref{tab:old_young_dust}.


   \begin{figure}[t!]
   \centering
   \vspace{-0.2cm}\includegraphics[width=0.5\textwidth]{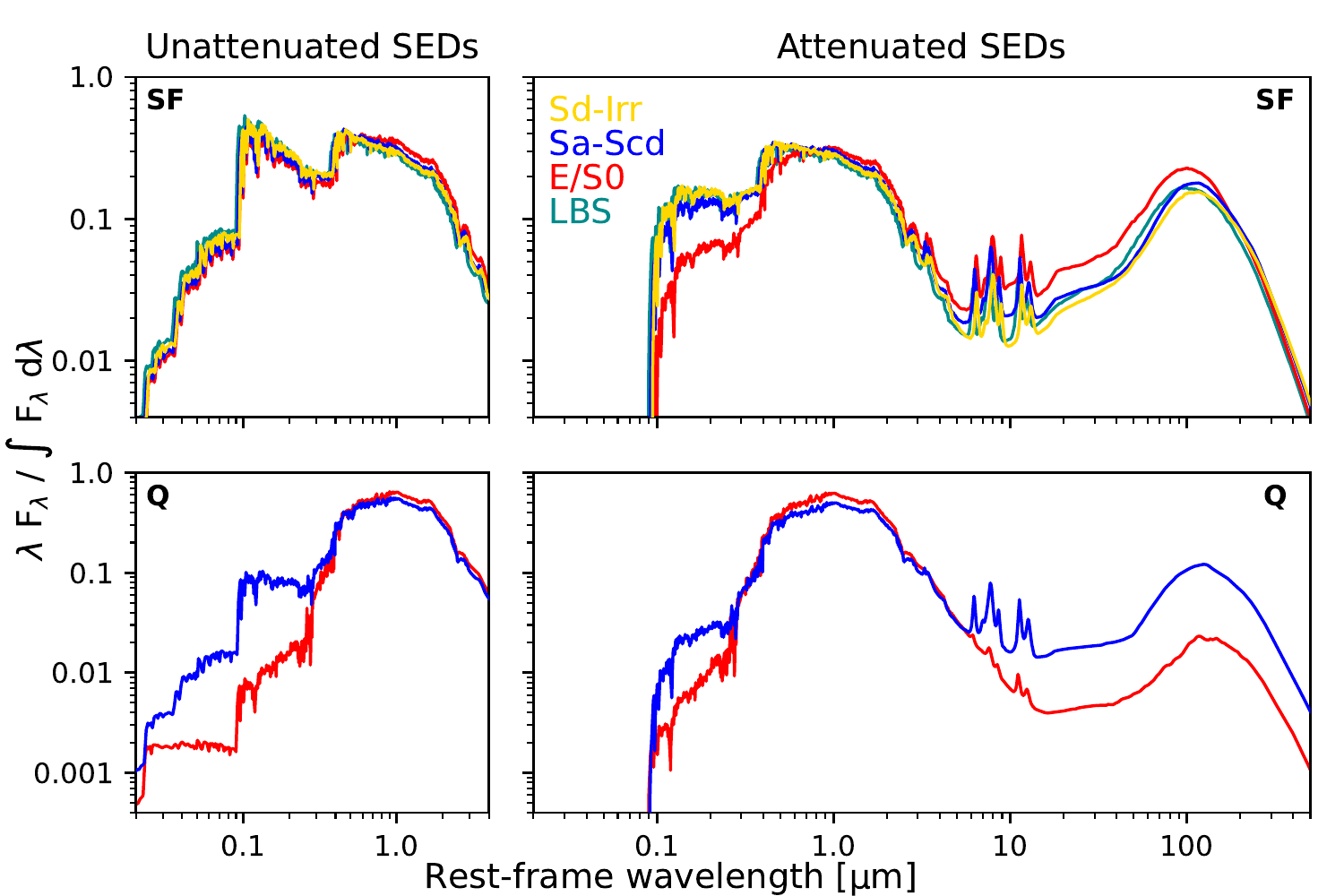}
   \caption{Median SEDs for the different morphological classes (different colours are for different classes, as indicated in the upper-right panel). The SEDs are grouped in SF galaxies (top panels) and Q galaxies (bottom panels). The unattenuated SEDs are plotted in the \textit{left panels}, while the attenuated ones are plotted in the \textit{right panels}.}
   \label{fig:SEDsMulti}
   \end{figure}

\section{Discussion}\label{discussion}
The coexistence of both types (Q and SF) in the various morphological types of galaxies is a very interesting finding and deserves a more thorough investigation in order to reveal the actual causes of this bimodal behaviour. In what follows we investigate possible differences between Q and SF galaxies regarding their energetic output (SEDs), their morphologies (as paramatrised by their $R_\text{e}$ and S\`ersic index), and the effects that the local environment may have on the galaxies.

\subsection{Differences between SEDs of Q and SF galaxies} \label{ap:stru}

In Fig.~\ref{fig:SEDsMulti} we compare the median attenuated and unattenuated SEDs (each individual SED normalised to its bolometric luminosity) of all morphological classes, but, grouped according to their star-forming activity. What is immediately evident is that the unattenuated SEDs of SF galaxies of all types of morphologies are very similar (almost identical; see top-left panel). There are only minor differences observed with the SEDs of SF LBS, SF Sa-Scd, and SF Sd-Irr galaxies, being slightly brighter, in the FUV-optical wavelength range ($\sim$0.1 dex at 0.2 $\mu$m), compared to those of SF E/S0 of the same bolometric luminosity. On the other hand, the infrared luminosity ($L_\text{IR}$) is also, on average, very similar between all types of SF galaxies with only SF E/S0s showing a higher peak value at $\sim 100 \mu$m (by $\sim$0.18 dex) compared to the rest morphologies. This suggests that all SF galaxies of the same bolometric luminosity, independently of their morphology, exhibit the same (within the statistical uncertainties imposed by our analysis) unattenuated stellar SEDs. When the effects of dust are taken into account, the differences become clearer in the attenuated stellar SEDs (top-right panel in Fig.~\ref{fig:SEDsMulti}) with SF LBSs, SF Sa-Scds, and SF Sd-Irrs showing similar SEDs which are, on average, brighter in the FUV/optical wavelengths, compared to those of SF E/S0 types (by $\sim$0.4 dex at 0.2 $\mu$m).

The Q galaxies (Sa-Scd, E/S0), on the other hand, show a diversity in the shape of their SEDs concerning both their attenuated and unattenuated stellar emission but also their dust emission (see the bottom panels in Fig.~\ref{fig:SEDsMulti}). Here we see that the unattenuated SEDs of the Q Sa-Scd galaxies, of the same bolometric luminosity, (bottom-left panel) show an enhanced stellar emission compared to those of Q E/S0s. This means that, on average, Q galaxies of the same bolometric luminosity in the Sa-Scd morphology bin are brighter in the FUV-optical wavelengths, compared to E/S0 types (meaning that they are more rich in young stars) but, at the same time, have more dust content (higher emission at FIR) leading to higher extinction values.

   \begin{figure*}[t!]
   \centering
   \includegraphics[width=0.86\textwidth]{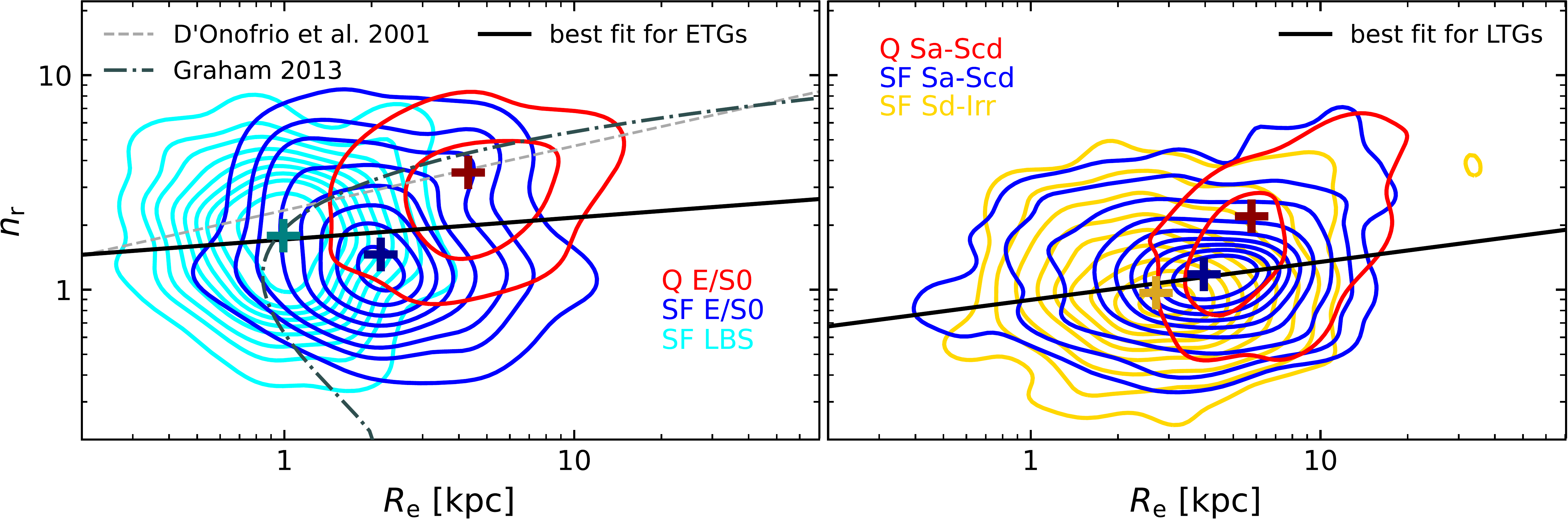}
   \caption{Kernel density estimates for the distribution of different kinds of galaxies in the $n_\text{r}$-$R_\text{e}$ plane. ETGs and LTGs are plotted in the \textit{left}, and \textit{right} panels, respectively. Different colours indicate different morphologies and star forming activities (as described in the insets in each panel) The median values of each subset are indicated by `+' symbols. The dashed-grey line and the dashed-dotted line are from  \citet{DOnofrio2001} and from \citet{Graham2013}, respectively (see the text for details).
   }
   \label{fig:contours}
   \end{figure*}

   \begin{table}
   \caption{Median r-band S\'{e}rsic indices ($n_\text{r}$) and effective radii ($R_\text{e}$) for the different galaxy populations.}
   \label{tab:structure}
   \begin{center}
   \begin{tabular}{l r c c}
\hline
\hline
   \multicolumn{1}{l}{Galaxy type} &
   \multicolumn{1}{r}{$N_\text{obj}$} &
   \multicolumn{1}{c}{$n_\text{r}$} &
   \multicolumn{1}{c}{$R_\text{e}$ [kpc]} \\
\hline
   Q E/S0 & 262 & 3.5 & 4.3 \\
   Q Sa-Scd & 52 & 2.2 & 5.8 \\
   SF LBS & 323 & 1.8 & 1.0 \\ 
   SF E/S0 & 231 & 1.5 & 2.2 \\
   SF Sa-Scd & 634 & 1.2 & 4.0 \\
   SF Sd-Irr & 692 & 1.0 & 2.7 \\
\hline
   \end{tabular}
   \end{center}
   \end{table}

\subsection{Structural characteristics} \label{structure}

The S\'{e}rsic index ($n_\lambda$) is considered a parameter to distinguish between different morphologies of galaxies (see, e.g. \citealt{Ravindranath2004}; \citealt{Vika2015}; \citealt{Mosenkov2019}, and references therein). In \citealt{Mosenkov2019}, it was shown that, although there is a lot of overlap, a borderline between LTGs and ETGs is at $n_\text{3.4}$=2 (with ETGs occupying the larger values). On the other hand, the effective radius ($R_\text{e}$) is a direct measure of the size of a galaxy, often relate to $n_\lambda$ (see, e.g., \citealt{DOnofrio2001}). 

Aiming at better understanding the morphological discrepancies between the star-forming and quiescent galaxies in our sample, we investigate how the structural parameters (in particular the r-band S\'{e}rsic index, $n_\text{r}$ and the effective radius, $R_\text{e}$) vary within galaxies of different morphologies and star-forming activity. As described in detail in \citet{Kelvin2012}, such structural parameters have been retrieved through single-S\'{e}rsic modelling using the SIGMA (Structural Investigation of Galaxies via Model Analysis) code. The galaxies in the GAMA sample are modelled with GALFIT \citep{GALFIT} (implemented in SIGMA) allowing for multiple parametric functions (e.g. S\'{e}rsic, exponential, Gaussian, etc.). The model of each galaxy is then produced with a downhill algorithm used by GALFIT  to minimise the global $\chi^2$. The final products are finally analysed by SIGMA for any obvious errors.

In Fig.~\ref{fig:contours} we plot Kernel density estimates of the $n_\text{r}$ as a function of the $R_\text{e}$. ETGs and LTGs are plotted in the left and right panels, respectively. The median values of $n_\text{r}$ and $R_\text{e}$, for all galaxy types are presented in Table~\ref{tab:structure}. For the ETGs (left panel in Fig.~\ref{fig:contours}) we see that there is an increase in n$_\text{r}$, on average, with increasing $R_\text{e}$, with SF LBS being least extended (median value of $R_\text{e}$=1~kpc), SF E/S0 being intermediate cases ($R_\text{e}$=2.2~kpc), and Q E/S0 being the most extended galaxies ($R_\text{e}$=4.3~kpc). The best-fit calculated for the current sample of ETGs is expressed by the equation
\begin{equation*}
    \text{log}(n_\text{r}) = 0.05^{+0.04}_{-0.04}~\text{log}(R_\text{e}[\text{kpc}]) + 0.25^{+0.02}_{-0.01},
\end{equation*}
with a correlation coefficient $\rho$ = 0.41, which is to be expected given the large dispersion of the distributions. Concerning the r-band S\'{e}rsic index the SF ETGs (LBS, E/S0) show, on average, lower indices (1.8, and 1.5, respectively) with the Q E/S0 showing a much steeper profile with $n_\text{r}=3.5$. Although it is clear that in all cases the median value of $n_\text{r}$ indicate non-exponential profiles ($n_\text{r}>1$), it is the Q E/S0 types that show clear cases of steep radial profiles. 
LTGs (right panel in Fig.~\ref{fig:contours}) show, on average, larger values of $R_\text{e}$, compared to ETGs, with SF Sd-Irr being the most compact LTGs (median value of $R_\text{e}$=2.7~kpc), followed by SF Sa-Scd ($R_\text{e}$=4~kpc) and Q Sa-Scd ($R_\text{e}$=5.8~kpc). The larger values, compared to ETGs, can be explained due to the more extended, disk-like, geometry of the spiral galaxies as well as the flattened and more irregular distribution of Sd-Irr types (as opposed to the generally more elliptical shape of ETGs). The increase of the $R_\text{e}$ as a function of the $n_\text{r}$ for the LTGs is described by the relation
\begin{equation*}
    \text{log}(n_\text{r}) = 0.21^{+0.01}_{-0.01}~\text{log}(R_\text{e}[\text{kpc}]) - 0.07^{+0.01}_{-0.01},
\end{equation*}
with a correlation coefficient $\rho$ = 0.28, which is again expected given the large dispersion of the distributions.

Concerning the S\'{e}rsic index LTGs show, on average lower values, compared to ETGs, consistent with exponential and more flattened profiles for SF Sa-Scd and SF Sd-Irr ($n_\text{r}=1.2$ and 1, respectively) with the Q Sa-Scd showing a steeper profile ($n_\text{r}=2.2$). The larger value of $n_\text{r}$ in the Q Sa-Scd galaxies may be an indication of a more prominent bulge component.

The fact that the median values of $n_\text{r}$ of SF LSB and E/S0 are closer to unity (see Table~\ref{tab:structure}) may indicate that, despite the overall elliptical shape of the galaxy, an embedded disk may be present.
\citet{Mahajan2018}, studying the Blue Spheroid (BSph) galaxies in the GAMA sample (here called the LBS), showed that although these galaxies have very similar structure to ellipticals, they resemble star-forming spirals in terms of age, metallicity and star formation. In their analysis they also revealed the underlying structures of the galaxies by fitting a S\`ersic profile. The residuals indicated the existence of a disk or a nuclear component in $\sim$38\% of BSph and in $\sim$43\% of ellipticals in the sample. In a different sample, consisting of 55 blue ETGs from the SDSS DR6, \citet{George2017} found that $\sim$58\% show similar structures, attributed to recent interactions. 

The star forming galaxies of all morphological types, show, systematically, lower median values of $n_\text{r}$ and $R_\text{e}$ in comparison to their Q counterparts. In addition, quiescent galaxies seem to be, on average, more extended than the star forming ones. 
This comes to an agreement with the study of \citet{Xu2022} where they find that for a sample of S0s in the SDSS-IV MaNGA survey, the SF S0s show lower, average, bulge S\'{e}rsic indices compared to a control sample of S0 normal galaxies. 
The general trend, despite the very large scatter, is that there is an increase of the value of $n_\text{r}$ with $R_\text{e}$. This is evident in both morphologies (ETGs and LTGs, left and right panels of Fig.~\ref{fig:contours}, respectively). 
Such a trend has previously been reported by \citet{DOnofrio2001}, for a volume- and magnitude-limited sample of 73 ETGs belonging to the Virgo and Fornax clusters (see the dashed-gray line in the left panel of Fig.~\ref{fig:contours}). 
This finding is also supported by the $n_\text{r}$-$R_\text{e}$ relation by \citet{Graham2013} (dash-dotted curve in the left panel of Fig.~\ref{fig:contours}) being approximately linear for high-$n$ ($n$>2) cases, but curved for the low-$n$ sources. Our data are in good agreement with this relation with the curved part of the fit mainly being shaped by the inclusion of LBS types.

\subsection{The role of the environment}

It has been found that the processes that control the star-forming activity in galaxies depend, not only on their stellar masses (see, e.g., Fig.~\ref{fig:m-s/morph}), but also on the environment the galaxies lie in and evolve (\citealt{Peng2010}; \citealt{Vulcani2010}; \citealt{Paccagnella2016}; \citealt{Darvish2017}). The dependence of SFR for a sample of GALEX NUV-detected SDSS ETGs with  environment was explored by \citet{Schawinski2009}. In the latter study the authors found that blue ETGs with high rates of star formation inhabit low-density environments. In another study, \citet{Goto2003b}, the authors concluded that there is a strong dependence of the star-formation activity with environment with the passive spiral galaxies mainly being cluster members. On the other hand, other studies downgrade the role of environment on the shaping of the properties of galaxies, making internal processes more efficient mechanisms. For instance, \citet{Rowlands2012} reported that the environment is not the only physical quantity affecting whether galaxies will evolve as passive, or not, with the processes that turn spirals to passive systems may happen in either high or low density environments. Furthermore, \citealt{Davies2019} found no or mild differences between the field and cluster galaxies in the DustPedia sample by studying their gas and dust properties.  

\begin{figure}[t!]
\centering
\includegraphics[width=0.5\textwidth]{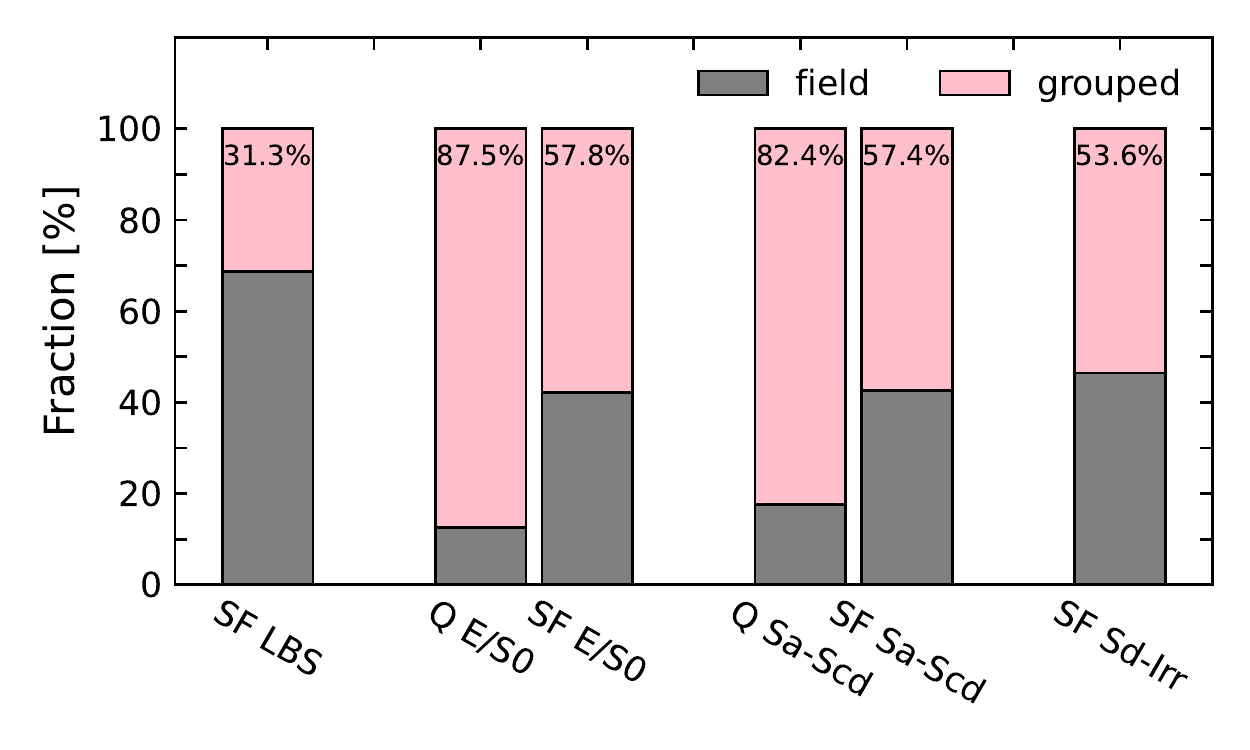}
\caption{The fractions of grouped (pink) and field (grey) galaxies for each subclass. The actual fraction of the grouped galaxies for each subclass is given in the top of each bar. 
}
\label{fig:bars-env}
\end{figure}

The GAMA Galaxy Group Catalogue (G$^3$C; \citealt{Robotham2011}) obtained by applying a friends-of-friends (FoF) algorithm provides information about the grouping of galaxies in the GAMA II equatorial regions of G09, G12 and G15 as well as the G02 survey region. By cross-matching the sample of the current study with the G$^3$Cv10 version of the catalogue, we find available information for 930 galaxies (42\% of the current sample). According to this classification, galaxy pairs or groups of more than two galaxy members are labelled as grouped, while field galaxies are totally isolated. The fractions of grouped and field galaxies in the current sample, for each morphological class, are presented in Fig.~\ref{fig:bars-env}. Out of the 930 galaxies, 491 are grouped, while 439 are isolated. Although the Q galaxies are under-represented (e.g., only 3 Q E/S0 out of 24 and 3 Q Sa-Scd out of 17) some general trends are evident. The general picture is that dense environments harbour more Q galaxies (85\% grouped, 15\% field Q galaxies) while SF galaxies are similarly divided between dense and isolated environments (51\% grouped, 49\% field SF galaxies). 
The only exception is for SF LBS galaxies with 69\% of these star-forming galaxies residing in isolated environments. The actual fractions of the field galaxies for each subsample are shown in Fig.~\ref{fig:bars-env}. Our findings are in good agreement with \citet{Fraser-McKelvie2018} who find that high-mass passive spiral galaxies reside mostly in groups. Similarly, \citet{Sampaio2022} studying cluster and field galaxies from the SDSS-DR7 found that 62\% of the cluster galaxies are 'red cloud' (Q) systems, while the majority (53\%) of the field galaxies are 'blue cloud' (SF) systems.

\begin{figure}[t!]
   \centering
   \includegraphics[width=0.5\textwidth]{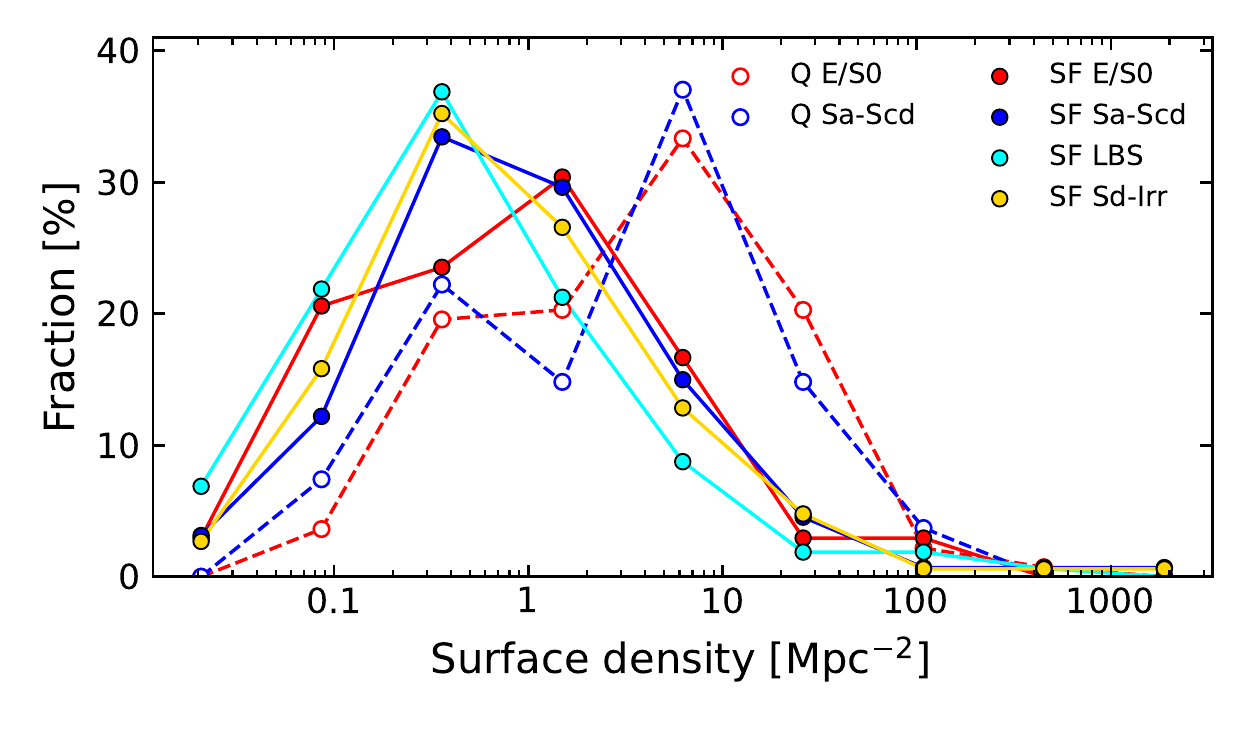}
   \caption{The morphology-density relation for the different types of galaxies. Open circles and dashed lines stand for the Q galaxies, while full circles and solid lines represent the SF galaxies. Since Q LBSs and Q Sd-Irrs consist of a small number of sources they are not presented in this plot. 
   }
   \label{fig:density}
   \end{figure}

A measure of the density of the local environment of a galaxy is the surface density, based on the distance to the 5th nearest neighbour. These measurements are provided for 1,049 sources in our sample in \citet{Brough2013}. In Fig.~\ref{fig:density} the data are grouped in 9 bins from 0.008 Mpc$^{-2}$ (minimum density) to 3,034 Mpc$^{-2}$ (maximum density) logarithmically spaced, with the lines connecting the measurements to guide the eye. From this plot it is evident that, as in Fig.~\ref{fig:bars-env}, SF galaxies tend to reside in less dense environments, as opposed to Q galaxies which tend to reside in rich environments. In particular, the peak of the distributions of all SF types of galaxies is around 0.3 Mpc$^{-2}$ (a median value of 0.49 Mpc$^{-2}$), while all Q types peak at around 6 Mpc$^{-2}$ (a median value of 2.71 Mpc$^{-2}$). It is notable, though that the range of the distributions is quite large ($\sim 0.1 - 10$ Mpc$^{-2}$ for SF types and $\sim 0.1 - 100$ Mpc$^{-2}$ for Q types) with significant overlap. More specifically, the median surface density of the environment of SF E/S0s and SF Sa-Scds is 0.67 Mpc$^{-2}$ and 0.60 Mpc$^{-2}$, respectively, while these values are 0.37 Mpc$^{-2}$ and 0.49 Mpc$^{-2}$ for SF LBS and SF Sd-Irrs, respectively. The Q galaxies, on the other hand, show higher median values of 2.79 Mpc$^{-2}$ for the Q E/S0s and 2.47 Mpc$^{-2}$ for the Q Sa-Scds, respectively.

\citet{Rowlands2012} found that both blue ETGs and passive spiral galaxies reside in environments with comparable densities to those of their `normal' counterparts, but probably this could be due to the small sample used (10 blue ETGs and 15 passive spirals). For a similar GAMA sample, such as the one used in this work, \citet{Pearson2021} found that the fraction of quiescent galaxies increases as the environment becomes more massive (supporting our findings). Our results also agree with the findings by \citet{Agius2015} who study 220 isolated ($\sim$0.1-10 Mpc$^{-2}$) ETGs from the GAMA/H-ATLAS sample \citep{Agius2013} and 33 ETGs from the HeViCS sample \citep{diSeregoAlighieri2013} belonging to the Virgo cluster ($\sim$25-500 Mpc$^{-2}$). They find that cluster galaxies have very little, or no, ongoing star-formation activity, while isolated galaxies present some ongoing star-formation, which, in some cases is comparable to that in star-forming spiral galaxies.

\subsection{Are SF Es a post-U/LIRG phase?}

In \citet{Paspaliaris2021} the authors investigated the evolution of the physical properties of luminous (10$^{11}$ L$_{\odot}$ $\leq$ $L_\text{IR}$ < 10$^{12}$ L$_{\odot}$; LIRGs) and ultra-luminous IR galaxies ($L_\text{IR}$ $\geq$ 10$^{12}$ L$_{\odot}$; ULIRGs) along the merging sequence. In an SED-fitting analysis, they studied 67 such local systems grouped in subsets according to their merging stage. The morphology of these systems was spanning from single (s) spiral galaxies or slightly disturbed spiral galaxies due to minor mergers (m) to spirals in a pre-merging stage (M1) and disturbed systems during a major merging event (M2-M5). As was indicated by the morphology of the late-merger (M5) systems in that study and also by earlier studies based on simulations (e.g. \citealt{Springel2005}, \citealt{Cox2006} \citealt{DiMatteo2008}, and \citealt{HopkinsP2013}), the product of the coalescence is an elliptical galaxy. Moreover, simulations have shown that, after major merging events, galaxies undergo a morphological change, prior to the drop in their SFR (\citealt{Dubois2016}; \citealt{Martin2018}; \citealt{Tacchella2019}; \citealt{Joshi2020}). The same result was found by \citet{Sa-Freitas2022} for a sample of blue E galaxies from SDSS DR12 \citep{Alam2015}. In the latter study, the blue Es although having enhanced SFR, they present a quenching behaviour attributed to a possible past merging event or rejuvenation process. In \citet{Paspaliaris2021}, however, no E galaxies (presumably being the final `relaxed' stage of a merging event) were analysed. This is a selection effect, since E galaxies mostly come with low luminosities (much less than the 10$^{11}$ L$_{\odot}$ LIRG limit). However, the diversity, in terms of morphology and star-forming activity, of the GAMAnear sample, analysed in this study, gives us the opportunity to make an investigation of the possibility of SF E galaxies being a post-U/LIRG product.

   \begin{figure}[t!]
   \centering
   \includegraphics[width=0.5\textwidth]{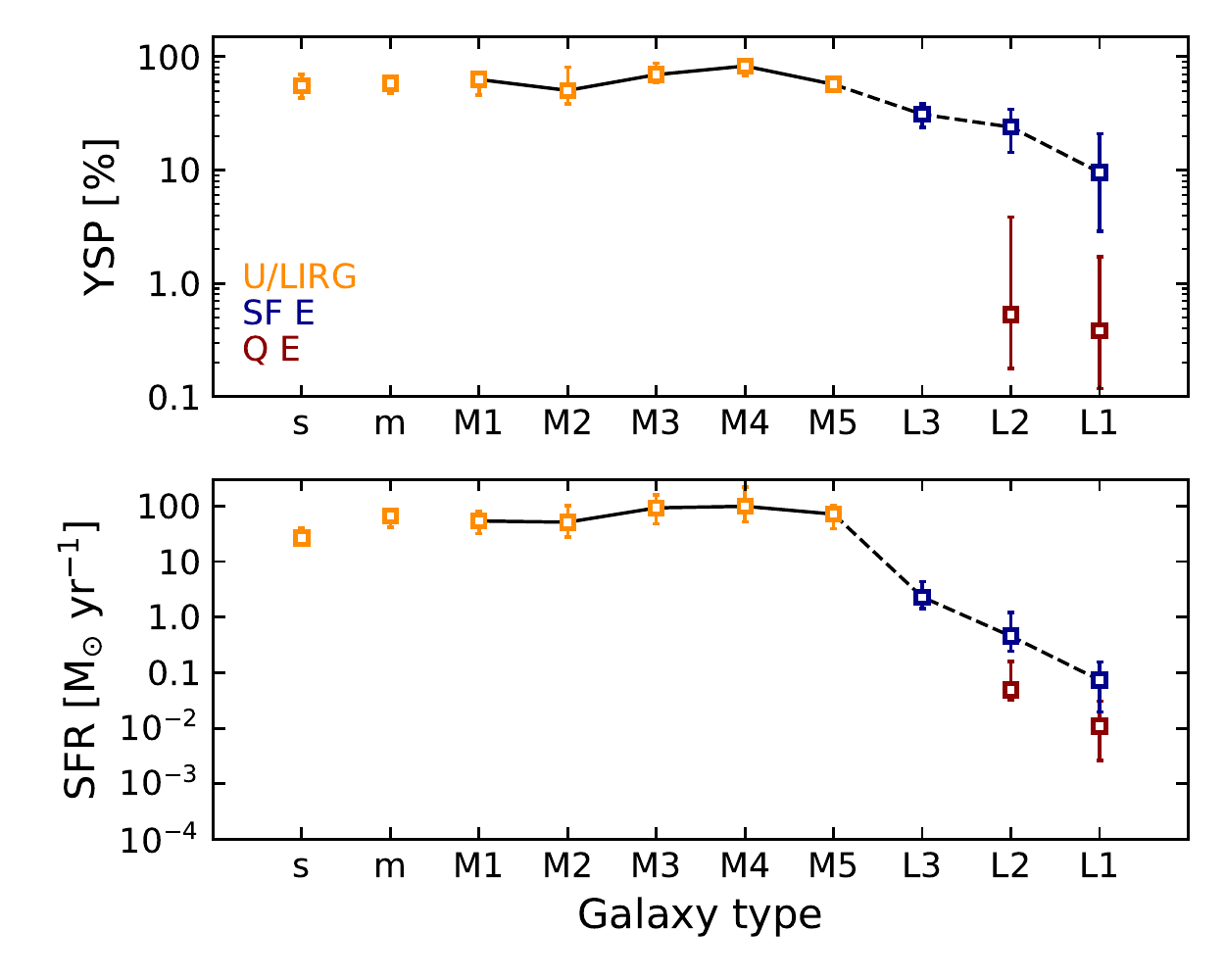}
   \caption{Evolution of the young stellar population (YSP; \textit{top panel}) and SFR (\textit{bottom panel}), along the U/LIRGs merging sequence and the post-U/LIRG phase. U/LIRGs from \citet{Paspaliaris2021} are shown in orange, while SF Es and Q Es are represented by blue and red squares, respectively. Error bars indicate the range between the 16th and 84th percentiles from the median. A black line connects the median values indicating the general trend.
   }
   \label{fig:evol}
   \end{figure}

Assuming that the IR luminosity of galaxies after the coalescence decreases with time, we grouped the SF Es in $L_\text{IR}$ bins, namely L1 ($L_\text{IR}$ < 10$^{9}$ L$_{\odot}$; 42 SF Es), L2 (10$^{9}$ L$_{\odot}$ $\leq$ $L_\text{IR}$ < 10$^{10}$ L$_{\odot}$; 69 SF Es), L3 ($L_\text{IR}$ $\geq$ 10$^{10}$ L$_{\odot}$; 23 SF Es) assuming that time evolves in the direction from L3 to L1 luminosity bins. We note that no E sources where found to have $L_\text{IR}$ $\geq$ 10$^{11}$ L$_{\odot}$. In the top panel of Fig.~\ref{fig:evol} we plot the median fractions of the young stellar population (YSP) for the different subsets of U/LIRGs and E galaxies. An increasing trend of the YSP in U/LIRGs (orange symbols) is observed between M2 and M4 systems (going from 50\% to 83\%), which then  decreases back to 57\% for M5 sources. For the SF Es (blue symbols) there is a decreasing trend with decreasing IR luminosity, however a jump of 0.4 dex exists between L2 and L1 sources. While L3 and L2 sources have YSP median values of 31\% and 24\% respectively, the corresponding value for L1 sources is 10\%. This difference could also indicate two different families of SF E galaxies, with the luminous ones (L2 and L3) consisting, mainly, of post-U/LIRG systems and L1 sources, with lower $L_\text{IR}$, might be galaxies rejuvenated by minor merging events or by accretion of gas available in their environment (see \citealt{Thom2012} for the latter scenario). The L1 and L2 bins of Q Es consist of 111 and 23 galaxies, respectively. We do not present the L3 bin for the Q Es because only one source was found within this $L_\text{IR}$ range. The YSP of Q Es (red symbols) also shows a decreasing trend, with decreasing IR luminosity. A notable characteristic that also supports the scenario which differentiates the L1 sources from the more luminous ones is that with increasing $L_\text{IR}$, the difference in the YSP between SF E and Q E subsets is larger (1.4 dex, 1.7 dex difference between L1, L2 Q/SF Es, respectively), while L3 bin is underrepresented for Q Es.

The evolution of the SFR is plotted in the bottom panel of Fig.~\ref{fig:evol}. In the case of U/LIRGs, a similar trend is found with the maximum occurring in M4 systems with a median value of 99 M$_{\odot}$ yr$^{-1}$. In M5 systems the median value is lower (71 M$_{\odot}$ yr$^{-1}$) and a dramatic decrease is observed in L3 and L2 SF Es (2.29 M$_{\odot}$ yr$^{-1}$ and 0.45 M$_{\odot}$ yr$^{-1}$, respectively). The decreasing trend of the SFR with decreasing $L_\text{IR}$ continues in L1 SF Es (0.07 M$_{\odot}$ yr$^{-1}$). The same trend is also observed for Q Es with 0.05 M$_{\odot}$ yr$^{-1}$ for L2 Q Es and 0.01 M$_{\odot}$ yr$^{-1}$ for L1 Q Es, respectively. A steep decrease in the SFR after the major merger has been also predicted by simulations. For instance, for realistic simulations of Milky Way-like galaxies at z=0, \citet{HopkinsP2013} finds the SFR falling to $\sim$2 M$_{\odot}$ yr$^{-1}$ in post-merging systems. In an earlier numerical study \citet{Springel2005} predicted that in the post-merger era, the SFR decreases to levels lower than 10 M$_{\odot}$ yr$^{-1}$, with similar results also reported by \citet{Cox2006} and \citet{DiMatteo2008}.
Deeper observations that could reveal `hidden' filamentary structures in the SF Es could confirm their possible merger-induced origin. As was also suggested by \citet{Rowlands2012} dusty ETGs, such as the SF Es in our sample, may be examples of post-merging systems with low inventories of hot gas, where a major star-formation event occurred a few Gyr before, creating their dust. This idea is also supported by the low on-average S\'{e}rsic indices found for the SF Es which could witness an imprint of a prior major merging event in these galaxies. Since x-ray emission traces the hot-gas content, such kind of observations are required to corroborate this scenario.

\section{Summary and conclusions}\label{sec:sum}

In this work we have examined a number of SED-derived global physical properties and the interplay between stars and dust of 2,209 GAMAnear galaxies observed by HSO. We classified the sources in subgroups according to their morphology and their dominant ionisation process (SF and Q galaxies). We excluded the sources that were flagged as AGN according to our classification method. By exploring the SFR-$M_\text{star}$ plane, we find that it is their star-formation activity (i.e., if they are classified as SF or Q) and not their morphological type that determines their place in the diagram. For instance, we find ETGs (E and S0) that are classified as SF and occupy the SFMS locus but also LTGs (Sa-Scd types) that are classified as Q and occupy the space below the SFMS. We also investigated if different local environments favour the existence of SF or Q galaxies of different morphologies. Furthermore, we explored if galaxies of different morphologies and star-formation activity show differences in their structural characteristics. Our main conclusions are as follows:

\begin{itemize}

    \item The median SEDs of galaxies of different star-formation activity indicate that, on average, SF galaxies, compared to their Q analogues, show enhanced dust emission with the dust being warmer.
    
    \item The place of a galaxy in the SFR-$M_\text{star}$ plane does not depend only on its morphology but mainly on its ability to convert gas into stars. In the GAMAnear sample, examined in this study, we find a large fraction (47\%) of the E/S0 types showing ongoing star-formation activity and 8\% of Sa-Scds being quiescent.
    
    \item SF E/S0s and SF LBS occupy the well-known `star-forming main-sequence' (SFMS), previously known to be, mainly, populated by LTGs. This finding as well as the fact that SF E/S0s and SF LBS show similar physical properties (e.g., dust-to-stellar-mass ratio and sSFR) and structural properties (S\'ersic index) indicate that LBSs may be smaller, in size, analogues of the SF ETGs.
    
    \item The low, average, dust-to-stellar-mass ratio found in Q Sa-Scds, compared to their SF counterparts, and their enhanced luminosity originating by the old stars indicate that their redder optical colours can be explained, not only due to dust deficiency but also due to enhanced old stellar population.
    
    \item The fraction of young stars in SF ETGs is found to be quite substantial (23\% and 25\% for SF E/S0 and SF LBS, respectively) with this fraction being only 1\% for the Q E/S0. SF LTGs, on the other hand, show similar fractions of young stars to that of SF ETGs (25\% for both types, SF Sa-Scds and SF Sd-Irrs) with this fraction being only 4\% for the Q Sa-Scds.  
    
    \item A significant contribution to the bolometric luminosity is found to be originating by dust ($f_\mathrm{abs}$) in SF ETGs (35\% in SF E/S0s and 19\% in LBSs) in contrast to that found for Q E/S0s (6\%). On the other hand, SF LTGs show high values of $f_\mathrm{abs}$ (27\% in SF Sa-Scd and 18\% in SF Sd-Irr) with the corresponding contribution by dust in Q Sa-Scds being 16\%.
    
    \item SF E/S0s use 24\% and 74\% of the luminosities originating from the old and the young stars to heat up the dust, while these fractions for the Q E/S0s are only 6\% and 36\%, respectively. SF Sa-Scds, on the other hand, use 16\% and 59\% of the luminosities originating from the old and the young stars to heat up the dust, while these fractions for the Q Sa-Scds are, surprisingly enough, very similar 12\% and 60\%, respectively. SF LBS and SF Sd-Irr galaxies show very similar fractions (11\% and 10\% of the luminosity of the old stars, and 44\% and 42\% of the luminosity of the young stars heating up the dust, respectively). 
    
    \item For the SF Sa-Scd, SF Sd-Irr, and the SF LBS types, the contribution of the young stars is, on average, the dominant source of dust heating (compared to the old stars) with values of 57\%, 59\% and 61\%, respectively. In Q Sa-Scd and SF E/S0s there is a significant contribution in the dust heating from the young stars (24\% and 47\%, respectively), while the dust in Q E/S0s is heated, almost exclusively, by old stars (91\%).
    
    \item The SED analysis conducted using CIGALE indicates that SF galaxies of the same bolometric luminosity, independently on their morphology, exhibit the same (within the statistical uncertainties) unattenuated stellar SED. Q galaxies, on the other hand, show different unattenuated SEDs.  
    
    \item Concerning the structural characteristics of the different types of galaxies we find that, for ETGs, there is an increase, on average, of the effective radius, with SF LBS being the less extended (median value of $R_\text{e}$=1~kpc), SF E/S0 being intermediate cases ($R_\text{e}$=2.2~kpc), and Q E/S0 being the most extended galaxies ($R_\text{e}$=4.3~kpc). LTGs, on the other hand, show, on average, larger values of $R_\text{e}$, compared to ETGs, with SF Sd-Irr being the most compact ones (median value of $R_\text{e}$=2.7~kpc), followed by SF Sa-Scd ($R_\text{e}$=4~kpc) and Q Sa-Scd ($R_\text{e}$=5.8~kpc). The r-band S\'{e}rsic index for SF ETGs (LBS, E/S0) show, on average, lower indices (1.8, and 1.5, respectively) with the Q E/S0 showing a much steeper profile with $n_\text{r}=3.5$. The fact that SF ETGs show relatively low values of the S\'{e}rsic index may indicate that, despite their overall elliptical shape, an embedded disk component may present in the centres of these galaxies. LTGs, on the other hand, show, on average, lower values, compared to ETGs, consistent with exponential and more flattened profiles for SF Sa-Scd and SF Sd-Irr ($n_\text{r}=1.2$ and 1, respectively) with the Q Sa-Scd showing a steeper profile ($n_\text{r}=2.2$).
    
    \item The local environment is found to affect star-forming activity in galaxies. We find that dense environments tend to favour the existence of quiescent galaxies in contrast to low-density environments which are mostly populated by SF galaxies.
    
\end{itemize}

Our analysis indicates that a significant fraction of ETGs may exhibit current star-formation activity with SFRs comparable to those of SF LTGs, while, on the opposite side many LTGs show ceased star-formation activity similar to quenched ETGs.
A number of the SF ETGs in our sample show signs of interactions or even host a disk-like or spiral-like structures at their centres. Similar indications have been shown by others (e.g., \citealt{Gomes2016b}, \citealt{George2017}, \citealt{Nyland2017}). \citet{Eales2017} proposed that the common hypothesis that ETGs are deficient in cold ISM is a consequence of the low instrumental sensitivity. Deep imaging (e.g. JWST, ELT or ALMA observations) might help reveal possible correlation between their enhanced SFRs and the morphological disturbances. Furthermore, X-ray and radio observations could shed light in the mechanisms that cease their star-forming activity in the quiescent LTGs, such as the ram pressure stripping in jellyfish galaxies. The results of the current work possibly indicate, however, that the star-formation activity and galaxy morphology transitions are not always linked needing different timescales to occur.


\begin{acknowledgements}
We would like to thank the anonymous referee for the useful comments and suggestions, which helped improving the quality of the manuscript.
EDP acknowledges the financial support from Greece and the European Union (European Social Fund - ESF) through the Operational Programme "Human Resources Development, Education and Lifelong Learning" in the context of the project "Strengthening Human Resources Potential via Doctorate research - Subaction 2: IKY grant program for doctoral candidates of Greek Universities" (MIS-5113934), implemented by the State Scholarships Foundation (IKY).
AN acknowledges the support of the Research Foundation - Flanders (FWO Vlaanderen). VAM acknowledges support by the Grant RTI2018-096686-B-C21 funded by MCIN/AEI/10.13039/501100011033 and by `ERDF A way of making Europe’. We would also like to thank Angel Ruiz (National Observatory of Athens) for constructive discussions regarding the UltraNest algorithm. This research made use of Astropy\footnote{\href{https://www.astropy.org}{http://www.astropy.org}}, a community-developed core Python package for Astronomy \citep{astropy13,astropy18}

GAMA is a joint European-Australasian project based around a spectroscopic campaign using the Anglo-Australian Telescope. The GAMA input catalogue is based on data taken from the Sloan Digital Sky Survey and the UKIRT Infrared Deep Sky Survey. Complementary imaging of the GAMA regions is being obtained by a number of independent survey programmes including GALEX MIS, VST KiDS, VISTA VIKING, WISE, Herschel-ATLAS, GMRT and ASKAP providing UV to radio coverage. GAMA is funded by the STFC (UK), the ARC (Australia), the AAO, and the participating institutions. VISTA VIKING data are based on observations made with ESO Telescopes at the La Silla Paranal Observatory under programme ID 179.A-2004.
\end{acknowledgements}

\bibliographystyle{aa}
\bibliography{References}

\appendix
\section{Validation of the results} \label{ap:validation}

\begin{figure}[t!]
   \centering
   \includegraphics[width=0.5\textwidth]{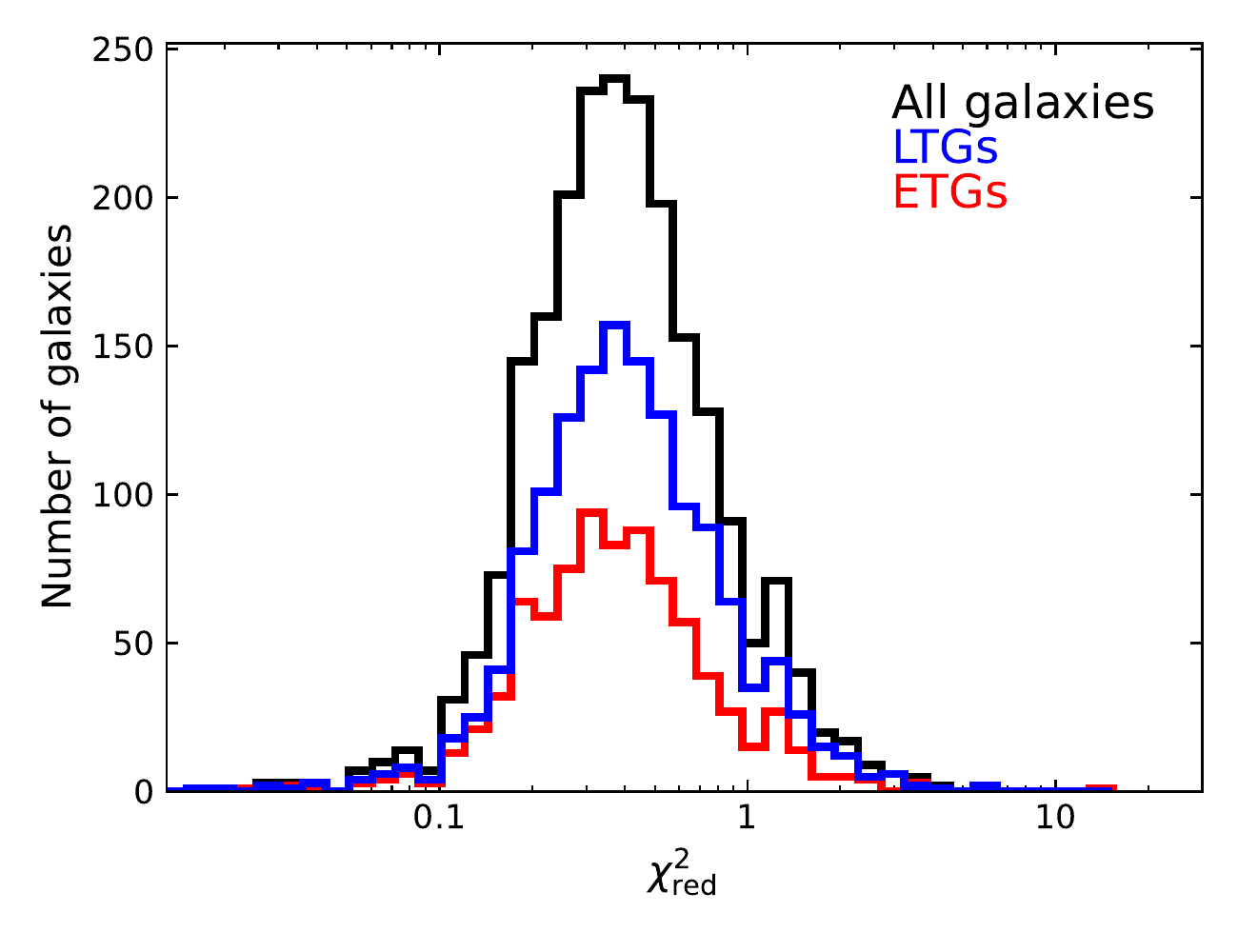}
   \caption{Distributions of the reduced $\chi^2$ for the 2,209 galaxies modelled with \texttt{CIGALE} (black line). The distributions of ETGs and LTGs are plotted in red and blue, respectively.
   }
   \label{fig:chi}
   \end{figure}

In order to investigate how accurately the derived parameters can be constrained from the multi-wavelength SED fitting that \texttt{CIGALE} performs, we examine the distribution of the reduced $\chi^2$ (estimated as the $\chi^2$ divided by the number of observations minus the population of the free parameters). The reduced $\chi^2$ distributions for different subsets of our sample are presented in Fig.~\ref{fig:chi}. The median reduced $\chi^2$ for all the 2,209 sources modelled is 0.38, while it is 0.36 for the ETGs and 0.40 for LTGs. Out of the sources modelled, 36 (1.6\%) have \mbox{$\chi^2_\text{red}$ > 2} and only 5 (0.2\%) have $\chi^2_\text{red}$ > 4, with the highest $\chi^2_\text{red}$ being equal to 15.3.

\subsection{Comparison between \texttt{CIGALE} and \texttt{MAGPHYS}}\label{app:comparison}

   \begin{figure*}[t!]
   \centering
   \includegraphics[width=\textwidth]{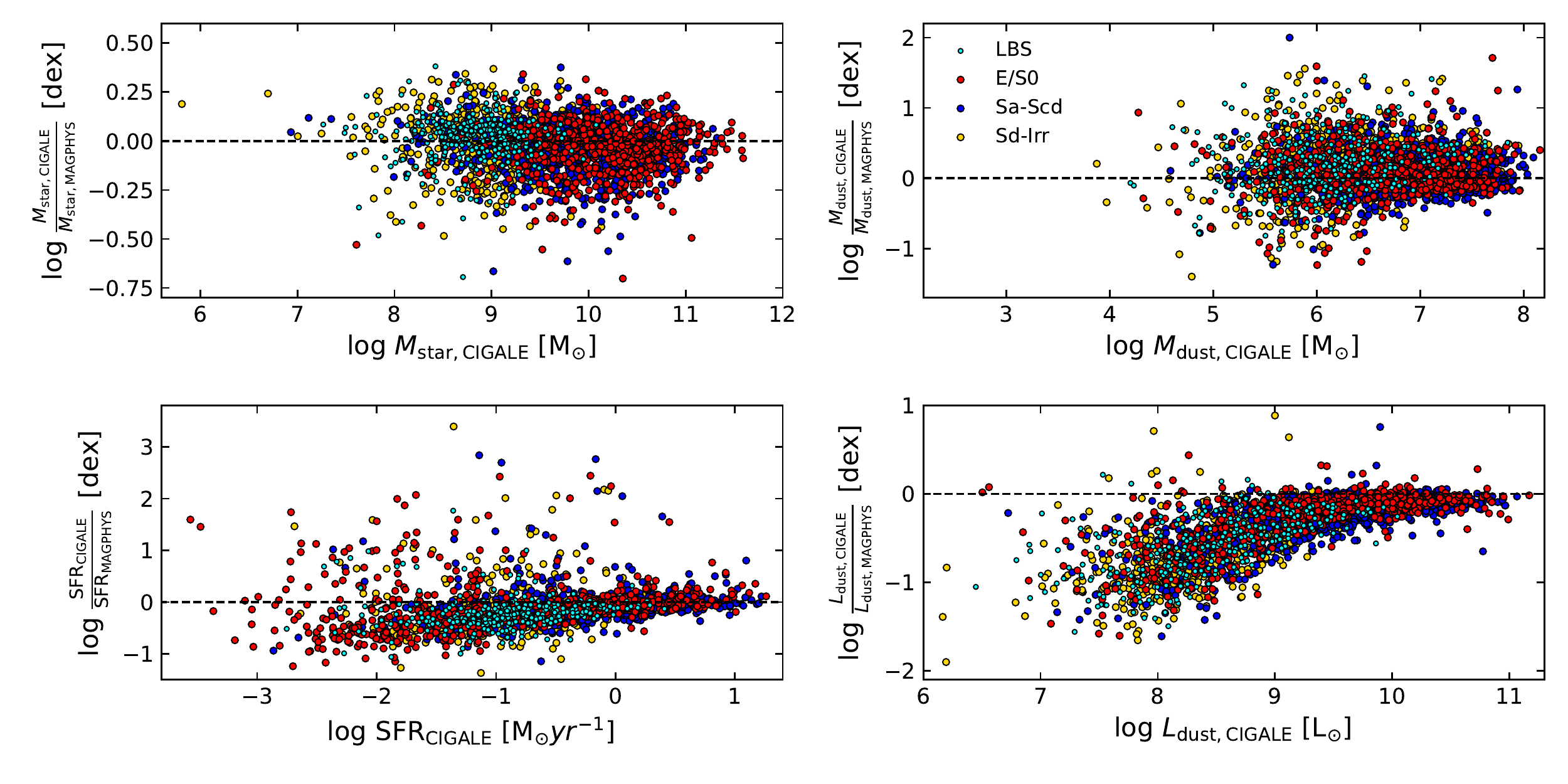}
   \caption{Comparison between the \texttt{CIGALE}-derived (this work) and the \texttt{MAGPHYS}-derived (\citealt{Driver2016}) properties of the galaxies in our sample. $M_\text{star}$, $M_\text{dust}$, SFR, and $L_\text{dust}$ are shown in the top-left, top-right, bottom-left, and bottom-right panels, respectively.}
   \label{fig:comp}
   \end{figure*}

The GAMA sample investigated in the current work has already been studied by \citet{Driver2016} and the basic parameters discussed in this paper have been computed using a similar approach with the SED-fitting code \texttt{MAGPHYS} \citep{daCunha2008}. In Fig.~\ref{fig:comp} we compare what is computed in the current work with what is provided in the GAMA survey DR3 file-server.

In the top-left panel of Fig.~\ref{fig:comp} we compare the stellar masses as derived by the two codes. Different colors indicate different morphological types as indicated in the top-right panel.
For the comparison of the stellar masses one needs to take into account that in \cite{daCunha2008} the \citet{Chabrier2003} IMF is considered, while the \citet{Salpeter} IMF is used in our analysis leading in a systematic underestimation of $M_\text{star}$ (see also \citealt{daCunha2010, Paspaliaris2021}).
To rescale stellar masses from Chabrier to Salpeter IMF, we divide by a constant factor of 0.61 as indicated in \citet{2014ARA&A..52..415M}. The stellar masses seem to be in a good agreement with roughly a $\pm$0.25 dex scatter and no obvious systematic offset (a median value of the differences of 0.0006 dex).

In the top-right panel of Fig.~\ref{fig:comp} the comparison of the dust masses as derived by \texttt{MAGPHYS} \citep{daCunha2008} and by \texttt{CIGALE} (current study) is presented. Compared to what is found for the stellar masses, the data show a larger dispersion, roughly $\pm$1 dex, with a median offset of 0.2 dex. This large dispersion and offset are to be expected since a different SED model and a different dust grain model are used in each fitting code. In \texttt{CIGALE} we use the \textsc{THEMIS} dust model. \textsc{THEMIS} is based on the optical properties of amorphous silicate and amorphous hydrocarbon materials measured in the laboratory \citep[see][and references therein]{Jones2017}, with a dust absorption coefficient $\kappa_{250{\mu}\text{m}}$ = 6.4 cm$^{2}$g$^{-1}$. On the other hand, as described in \citet{daCunha2008} \texttt{MAGPHYS} models the IR emission from stellar birth-clouds as the sum of a component of polycyclic aromatic hydrocarbons (PAHs), a mid-IR continuum component originating from hot grains (130-250 K) and a component of grains in thermal equilibrium (adjustable temperature 30-60 K). In addition, for the emission from the ISM it uses the latter three components to reproduce the spectral shape of the Milky Way diffuse cirrus emission along with a cold grain component in thermal equilibrium (adjustable temperature 15-25 K). The warm and cold components are described by modified blackbody spectra and absorption cross section with spectral indices $\beta$ equal to 1.5 and 2.0, respectively, normalised to $\kappa_{850{\mu}\text{m}}$ = 0.77 cm$^{2}$g$^{-1}$ (\citealt{Dunne2000}; \citealt{James2002}).

The comparison of the SFR, as estimated by the two codes, is presented in the bottom-left panel of Fig.~\ref{fig:comp}. In order to account for the different IMFs used between the two codes, we use a constant conversion factor of 0.63 to convert from the Chabrier SFRs to Salpeter SFRs (see \citealt{2014ARA&A..52..415M}). A systematic \texttt{CIGALE} underestimation of SFR is observed for sources with SFR less than about one M$_{\odot}\text{yr}^{-1}$. This might be explained by the different SFH modules used in the two codes. In particular, the SFH used in the current study is a late burst or quench superimposed to a more passively evolving component, while the one used in \texttt{MAGPHYS} is a continuous exponentially declining SFH, with additional random bursts of star formation.

\begin{figure}[t!]
   \centering
   \includegraphics[width=0.5\textwidth]{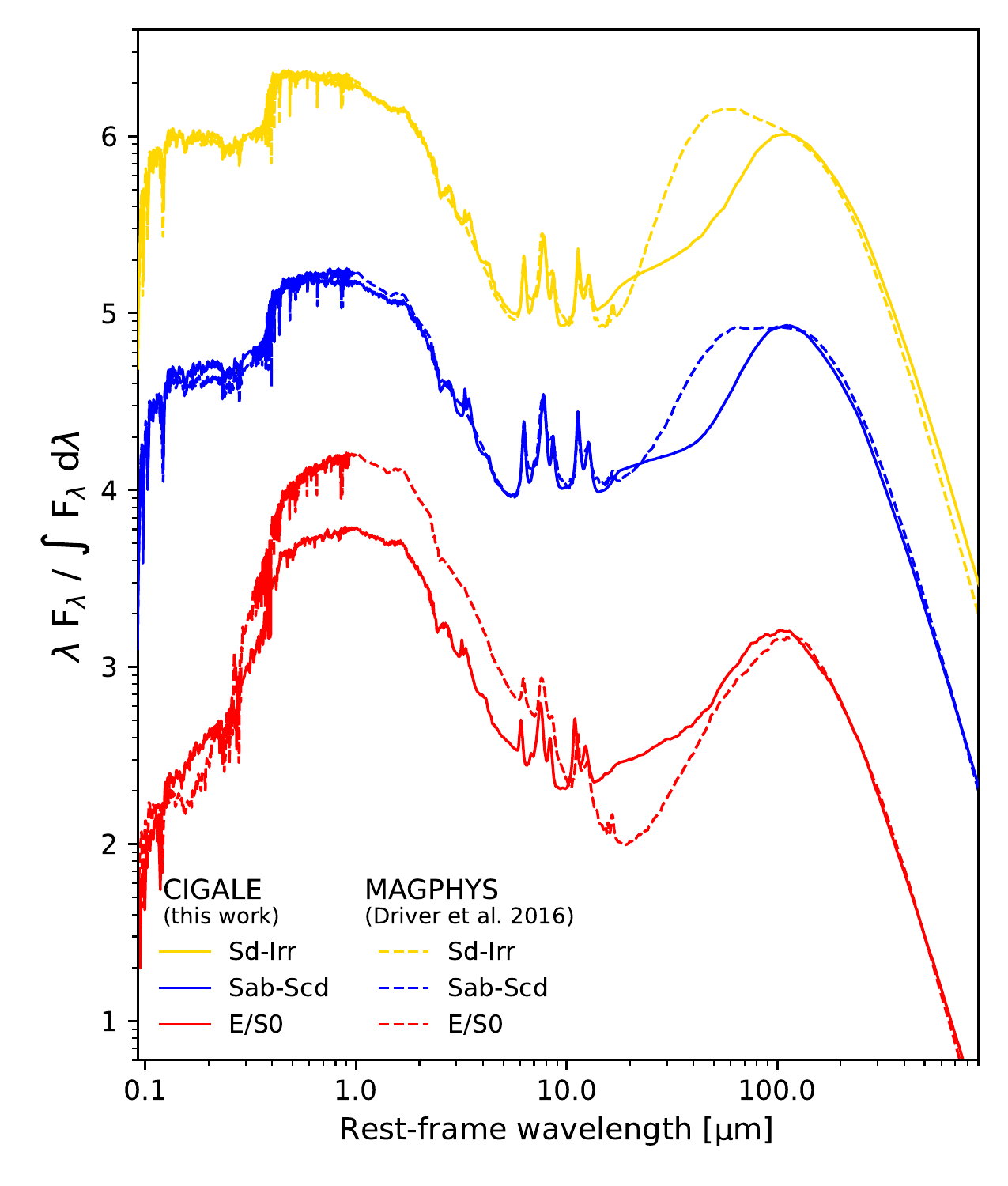}
   \caption{Comparison of the median SEDs fitted by \texttt{CIGALE} in the current work (solid curves) and by \texttt{MAGPHYS} (dashed curves) presented in \citet{Driver2016}.
   }
   \label{fig:compSEDs}
   \end{figure}

In the case of dust luminosity (bottom-right panel) there is clear underestimation of $L_\text{dust}$, as computed by \texttt{CIGALE}, especially evident at low luminosities. A tendency of the \texttt{MAGPHYS} code to maximise the dust content within the bounds defined by the errors of the FIR observations has already been reported by \citet{Driver2011}. Several other studies (e.g. \citealt{Pappalardo2016}; \citealt{Bianchi2018}; \citealt{Hunt2019}) have also reported a broader bump at the FIR emission, when \texttt{MAGPHYS} is used, especially in the case where insufficient wavelength coverage exists in the range of 24-100 $\mu$m. This effect is clearly observed in SEDs of Fig.~\ref{fig:compSEDs}, where the median templates fitted in this work using \texttt{CIGALE} (solid curves) are compared with the ones produced by \citet{Driver2016} with \texttt{MAGPHYS} (dashed lines). While the SEDs of Sab-Scd (blue) and Sd-Irr galaxies (yellow) are almost identical up to the $\sim$10 $\mu$m regime, the FIR bump is broader in the \texttt{MAGPHYS} SEDs and peaks at shorter wavelengths. Thus the overestimation of the dust luminosity by \texttt{MAGPHYS} is expected, since in both cases it is calculated by integrating the SEDs in the area of 8-1000 $\mu$m.

\end{document}